\documentclass[a4paper, 11pt]{article}

\usepackage[english]{babel}
\usepackage{amsmath, amsthm, amssymb, amsfonts, bm}
\usepackage{mathrsfs}
\usepackage{bbm}
\usepackage{float}
\usepackage{color,graphicx,epstopdf}
\usepackage{tikz,pgf}
\usetikzlibrary{arrows,shapes,shadows,petri}
\usepackage{dsfont}
\usepackage{hyperref}
\usepackage{harvard}
\citationstyle{dcu}
\usepackage{indentfirst}
\usepackage{subcaption}

\DeclareMathOperator{\ULG}{U}

\DeclareMathOperator{\so}{\mathfrak{so}}
\DeclareMathOperator{\su}{\mathfrak{su}}
\DeclareMathOperator{\spla}{\mathfrak{sp}}
\DeclareMathOperator{\ula}{\mathfrak{u}}

\newcommand{\R}{\ensuremath{\mathbb{R}}}
\newcommand{\C}{\ensuremath{\mathbb{C}}}
\newcommand{\QQ}{\ensuremath{\mathcal{Q}}}
\newcommand{\HH}{\ensuremath{\mathcal{H}}}
\newcommand{\Hn}{\ensuremath{\mathcal{H}^{\otimes n}}}
\def\P{\ensuremath{\mathsf{P}}}
\newcommand{\PP}{\ensuremath{\mathpzc{P}}}
\newcommand{\A}{\ensuremath{\mathsf{A}}}
\newcommand{\B}{\ensuremath{\mathsf{B}}}
\newcommand{\Q}{\ensuremath{\mathsf{Q}}}
\newcommand{\M}{\ensuremath{\mathsf{M}}}
\newcommand{\U}{\ensuremath{\mathsf{U}}}
\newcommand{\X}{\ensuremath{\mathsf{X}}}
\newcommand{\oH}{\ensuremath{\mathsf{H}}}
\newcommand{\uu}{\ensuremath{\mathsf{u}}}
\newcommand{\bra}[1]{\ensuremath{|#1\rangle}}
\newcommand{\ket}[1]{\ensuremath{\langle#1|}}
\newcommand{\braket}[2]{\ensuremath{|#1\rangle\langle#2|}}

\newcommand{\rmp}{\ensuremath{\mathrm{p}}}

\newcommand{\bluetext}[1]{{\color{blue}#1}}

\newcommand*\mcapinn[2]{\vcenter{\hbox{$\mathsurround=0pt
  \ifx\displaystyle#1\textstyle\else#1\fi\bigcap$}}}

\newcommand*\mcupinn[2]{\vcenter{\hbox{$\mathsurround=0pt
  \ifx\displaystyle#1\textstyle\else#1\fi\bigcup$}}}

\DeclareFontFamily{OT1}{pzc}{}
\DeclareFontShape{OT1}{pzc}{m}{it}{<-> s * [1.200] pzcmi7t}{}
\DeclareMathAlphabet{\mathpzc}{OT1}{pzc}{m}{it}

\graphicspath{{./images/}}

\newtheorem{theorem}{Theorem}
\newtheorem{definition}{Definition}

\newtheorem{lemma}{Lemma}
\newtheorem{remark}{Remark}
\newtheorem{proposition}{Proposition}
\newtheorem{assumption}{Assumption}
\newtheorem{example}{Example}

\renewcommand{\bluetext}[1]{#1}

\begin{document}

\title{\bf Measurement-Induced Boolean Dynamics and Controllability  for Quantum Networks}
	
\author{Hongsheng Qi\thanks{Key Laboratory of Systems and Control, Institute of Systems Science, Academy of Mathematics and Systems Science, Chinese Academy of Sciences, Beijing 100190, China; Research School of Electrical, Energy and Materials Engineering,  The Australian National University, Canberra 0200, Australia. E-mail: qihongsh@amss.ac.cn.}, Biqiang Mu\thanks{Key Laboratory of Systems and Control, Institute of Systems Science, Academy of Mathematics and Systems Science, Chinese Academy of Sciences, Beijing 100190, China}, Ian R. Petersen\thanks{Research School of Electrical, Energy and Materials Engineering,  The Australian National University, Canberra 0200, Australia. E-mail: ian.petersen@anu.edu.au.}, Guodong Shi\thanks{Australian Center for Field Robotics, School of Aerospace, Mechanical and Mechatronic Engineering, The University of Sydney, NSW 2008, Australia. E-mail: guodong.shi@sydney.edu.au.}}

\date{}

\maketitle

\begin{abstract}                          
In this paper, we study dynamical quantum networks which evolve according to Schr\"odinger equations but subject to sequential local or global quantum  measurements. A network of qubits forms a composite quantum system whose state undergoes unitary evolution in between periodic measurements, leading to hybrid quantum dynamics with random jumps at discrete time instances along a continuous orbit. The measurements either act on the entire network of qubits, or only a subset of qubits. First of all, we reveal that this type of hybrid quantum dynamics induces   probabilistic Boolean recursions  representing the measurement outcomes. With global measurements, it is shown that  such resulting Boolean recursions define Markov chains whose state-transitions are fully determined by the network Hamiltonian and the measurement observables. Particularly, we establish an explicit and  algebraic representation of the underlying recursive random mapping driving   such induced  Markov chains. Next, with local measurements, the resulting probabilistic Boolean dynamics is shown to be no longer Markovian. The state transition probability at any given time becomes dependent on the entire history of the sample path, for which we establish a recursive way of computing such non-Markovian  probability transitions. Finally, we adopt the classical bilinear control model for the continuous  Schr\"odinger evolution, and show how the measurements affect the controllability of the quantum networks. 
\end{abstract}


\section{Introduction}
Quantum systems admit drastically different  behaviors compared to classical systems in terms of state representations, evolutions, and measurements, based on which there holds the promise to develop  fundamentally new computing and  cryptography infrastructures for our society \cite{Nielsen}. Quantum states are described by vectors in finite or infinite dimensional  Hilbert spaces; isolated quantum systems exhibit   closed dynamics described by Schr\"odinger equations; performing measurements over a quantum system yields  random outcomes and creates back action to the system being measured. When interacting with    environments, quantum systems admit more complex  evolutions which are often approximated by various  types of master equations. The study of  the evolution   and manipulation of quantum states has been one of the central problems in the fields of quantum science and engineering \cite{Altafini2012}. 

For the control or manipulation of quantum systems, we can carry out feedforward control by directly revising the Hamiltonians in the Schr\"odinger equations \cite{bro72}, resulting in  bilinear control systems. Celebrated results have been established regarding the controllability of such systems from the perspective of geometric nonlinear control \cite{jur72,bro72,bro00,Schirmer2001,alb03,Li2009,belabbas2018}. In the presence of external environments, one can also directly engineer the interaction between the quantum system of interest and the environments, e.g.,  \cite{Wang2010,ticozzi2010}.  Feedforward can also be carried out by designing a sequence of measurements from different bases \cite{Rabitz-PRA}, where the quantum back actions from the measurements  are utilized as a control mean.  

Feedback control can also be carried out for quantum systems via  coherent feedback \cite{ian2008} or measurement feedback  \cite{Belavkin1979,QubitFeedback2014}. In coherent feedback, the outputs of a quantum system are fed back to the control of the inherent or interacting Hamiltonians. While in measurement feedback, the measurement outcomes are fed back to the selection of the future measurement bases. Introducing feedback to the control of quantum systems on one hand improves the robustness of the closed-loop system, and on the other hand, the resulting quantum back actions intrinsically  perturb  the system states subject to the quantum uncertainty principle.

Qubits, the so-called quantum bits, are the simplest quantum states with a two-dimensional state space.  \bluetext{Qubits naturally form networks in various forms of interactions:  they can  interact directly with each other by coupling Hamiltonians  in a quantum composite system~\cite{Altafini2002}; implicitly  through coupling with local environments~\cite{Shi-TAC}; or through local quantum operations such as measurements and classical communications on the operation outcomes~\cite{nature2010}.  } Qubit networks have become canonical models for quantum mechanical states and interactions between particles and fields under the notion of spin networks   \cite{Yamamoto2014}, and for   quantum information processing platforms in computing and communication \cite{nature2010,shi-sr}.  The control of qubit networks has been studied in various forms \cite{alb02,Wang2012,dir08,Shi-TAC,li2017}. 

In this paper, we study dynamical qubit  networks which evolve as a collective  isolated quantum system but  subject to sequential local or global measurements. Global measurements are represented by observables applied to all qubits in the network, and local measurements only apply to a subset of qubits and therefore the state information of the remaining qubits becomes hidden.  We reveal  that this type of hybrid quantum dynamics   induces   probabilistic Boolean recursions  representing  the   measurement outcomes, defining a quantum-induced probabilistic Boolean network. Boolean networks, introduced by Kauffman in the 1960s \cite{Kauffman1969} and then extended to 
probabilistic Boolean networks \cite{Probabilistic-Boolean-Network}, have been a classical model for gene regulatory interactions. The behaviors of Boolean dynamics are quite different compared to classical dynamical systems described by differential or difference equations due to their combinatorial natures, and their studies have been focused on the analytical or approximate characterizations to the steady-state orbits and controllability   \cite{Chaves2013,Cheng2009}. 
The contributions of the paper are summarized as follows: 

 \bluetext{
	\begin{itemize}
		\item  Under global  measurements, the induced Boolean recursions   define Markov chains  for which we establish a purely   algebraic representation of the underlying  recursive random mapping. The representation is in the form of random linear systems embedded in a high dimensional  real space. 
		
		\item Under local measurements,  the resulting probabilistic Boolean dynamics is no longer Markovian. The transition probability at any given time relies  on the entire history of the sample path,  for which we establish a recursive computation scheme.  
		
		\item In view of  the classical bilinear model for closed quantum systems, we demonstrate  how the measurements affect the controllability of the quantum networks. In particular, we show that practical quantum state controllability is already enough to guarantee almost sure Boolean state controllability.  
	\end{itemize}
}

The remainder of the paper is organized as follows. Section \ref{secpre} presents a collection of preliminary knowledge and theories which are essential for our discussion. Section \ref{secmodel} presents the qubit network model for the study. Section \ref{secboolean} focuses on the induced Boolean network dynamics from the measurements of the dynamical qubit network. Section \ref{seccontrollability} then turns to the controllability of such qubit networks under bilinear control. Finally Section \ref{seccon} concludes the paper with a few remarks.

\section{Preliminaries}\label{secpre}

In this section, we present some preliminary knowledge on quantum system states and measurements, quantum state evolution and bilinear control, probabilistic Boolean networks, and Lie algebra and groups, in order to facilitate a self-contained presentation.

\subsection{Quantum States and Projective Measurements}

The state space of any isolated quantum system is a complex vector space with inner product, i.e., a Hilbert space $\HH_N\simeq\C^N$ for some integer $N\ge 2$.
The system state is described by a unit vector in $\HH_N$ 
denoted by $\bra{\varphi}$, where $\bra{\cdot}$ is known as the Dirac notion for vectors representing quantum states. The complex conjugate transpose of $\bra{\varphi}$ is denoted by $\ket{\varphi}$.
One primary feature that distinguishes quantum systems from classical systems is the state space of composite system consisting of one or more subsystems. The state space of a composite quantum system is the tensor product  of the state space of each component system. 
As a result, the states of a composite quantum system of two subsystems with state space $\HH_A$ and $\HH_B$, respectively, are complex linear combinations of $\bra{\varphi_A}\otimes \bra{\varphi_B}$, where $\bra{\varphi_A}\in\HH_A$, $\bra{\varphi_B}\in\HH_B$.

Let $\mathscr{L}_*(\HH_N)$ be the space of linear operators over $\HH_N$. For a quantum system associated with state space $\HH_N$, a projective measurement is described by an observable $\M$, which is a Hermitian operator in $\mathscr{L}_*(\HH_N)$. The observable $\M$ has a spectral decomposition in the form of
$$
\M=\sum\limits_{m=0}^{N-1} \lambda_m\P_m,
$$
where $\P_m$ is the projector onto the eigenspace of $\M$ with eigenvalue $\lambda_m$. The possible outcomes of the measurement correspond to the eigenvalues  $\lambda_m$, $m=0,\dots,N-1$ of the observable. Upon measuring the state $\bra{\varphi}$, the probability of getting result $\lambda_m$ is given by
$
p(\lambda_m)=\ket{\varphi}\P_m\bra{\varphi}
$.
Given that outcome $\lambda_m$ occured, the state of the quantum system immediately after the measurement is
$\frac{\P_m\bra{\varphi}}{\sqrt{p(m)}}$.

\subsection{Closed Quantum Systems}

The time evolution of the state $\bra{\varphi(s)}\in\HH_N$ of a closed quantum system is described by a Schr\"odinger equation:
\begin{align}\label{eq:sch}
	\bra{\dot\varphi(s)} = -\imath \oH(s) \bra{\varphi(s)},
\end{align}
where $\oH(s)$ is a Hermitian operator over $\HH_N$ known as the Hamiltonian of the system at time $s$. 
Hamiltonians relate to physical quantities such as momentum, energy etc. for quantum systems.
Here without loss of generality the initial time is assumed to be $s=0$.
For any time instants $s_1,s_2\in[0,\infty)$, there exists a unique unitary operator $\U_{[s_1,s_2]}$ such that
\begin{align}
	\bra{\varphi(s_2)}=\U_{[s_1,s_2]}\bra{\varphi(s_1)}
\end{align}
along the Schr\"odinger equation \eqref{eq:sch}.

\subsection{Bilinear Model for Quantum Control}

Let $\mathscr{O}_*(\HH_N)$ be the space of Hermitian operators over $\HH_N$. The basic bilinear model for the control of a quantum system is defined by letting $\oH(s)=\oH_0+\sum_{\ell=1}^p u_\ell(s)\oH_\ell$ in the Schr\"odinger equation \eqref{eq:sch}, where $\oH_0\in\mathscr{O}_*(\HH_N)$ is the unperturbed or internal Hamiltionian, and $\oH_\ell\in\mathscr{O}_*(\HH_N)$, $\ell=1,\dots,p$ are the controlled Hamiltonians with the $u_{\ell}(s),\ell=1,\dots,p$ being control signals as real scalar functions. This leads to
\begin{align}\label{eq:sch-control}
	\begin{aligned}
		\bra{\dot\varphi(s)} =& -\imath \left(\oH_0+\sum_{\ell=1}^p u_\ell(s)\oH_\ell\right) \bra{\varphi(s)}\\
		:=& \left(\A+\sum_{\ell=1}^p u_\ell(s)\B_\ell\right) \bra{\varphi(s)},
	\end{aligned}
\end{align}
where $\A=-\imath\oH_0$, and $\B_\ell=-\imath\oH_\ell$.
The background of this model lies in physical quantum systems for which we can manipulate their Hamiltonians.
Let $\X(s)$ be the operator defined for $s\in[0,\infty)$ satisfying 
\begin{align}\label{eq:propogator}
	\bra{\varphi(s)} = \X(s)\bra{\varphi(0)}
\end{align}
for all $s\ge 0$ along the equation \eqref{eq:sch-control}.
It can be shown that the evolution matrix operator $\X(s)$ is described by
\begin{align}\label{eq:propogator-final}
	\dot{\X}(s)=\left(\A+\sum_{\ell=1}^p u_\ell(s)\B_\ell\right)\X(s)
\end{align}
starting from $\X(0)=\mathrm{I}_{N}$.

The following two definitions specify basic controllability questions arising from the bilinear model \eqref{eq:sch-control}.

\begin{definition}
	\label{def:psc}
	The system \eqref{eq:sch-control} is pure state controllable if for every pair quantum states $\bra{\varphi_0},\bra{\varphi_1}\in\HH_N$, there exist $\mu>0$  and control signals $u_1(s),\dots,u_p(s)$ for $s\in[0,\mu]$ such that the solution of \eqref{eq:sch-control} yields $\bra{\varphi(\mu)}=\bra{\varphi_1}$ starting from $\bra{\varphi(0)}=\bra{\varphi_0}$.
\end{definition}

\begin{definition}
	\label{def:esc}
	The system \eqref{eq:sch-control} is equivalent state controllable if for every pair quantum states $\bra{\varphi_0},\bra{\varphi_1}\in\HH_N$, there exist $\mu>0$,  control signals $u_1(s),\dots,u_p(s)$ for $s\in[0,\mu]$, and a phase factor $\phi$  such that the solution of \eqref{eq:sch-control} yields $\bra{\varphi(\mu)}=e^{\imath\phi}\bra{\varphi_1}$ starting from $\bra{\varphi(0)}=\bra{\varphi_0}$.
\end{definition}

\begin{remark}
	From a physical point of view, the states $e^{\imath\phi}\bra{\varphi}$ and $\bra{\varphi}$ are the same as the phase factor $e^{\imath\phi}$ contributes to no observable effect.
\end{remark}

\subsection{Probabilistic Boolean Networks}

A Boolean network consists of $n$ nodes in $\mathrm{V}=\{1,2,\dots,n\}$ with each node $i$ holding a logical value $x_i(t)\in\{0,1\}$ at discretized time $t=0,1,2,\cdots$. Denote $\mathbf{x}(t)=\big[x_1(t)\dots x_n(t)\big]$, and let $\mathcal{S}$  denote the space containing all functions that map $\{0,1\}^{n}$ to $\{0,1\}^{n}$.
The evolution of the network states $\mathbf{x}(t)$ can then be described by the functions in $\mathcal{S}$. In a probabilistic Boolean network, at each time $t=0,1,2,\dots$, a   function $ f_t$ is drawn randomly  from $\mathcal{S}$ according to some underlying distributions, and the network state evolves according to
\begin{align}
	\mathbf{x}(t+1)= f_t\big( \mathbf{x}(t)\big).
\end{align}
To be precise,  $\Omega=\mathcal{S}\times \mathcal{S}\times \cdots$  and $
\mathcal{F}=2^\mathcal{S}\times 2^\mathcal{S}\times \cdots$
are the overall sample space and event algebra $\mathcal{F}$ equipped with probability measure  $\mathbb{P}$, where $\omega=(\omega_0,\omega_1,\dots)\in \Omega$. Let $\mathcal{F}_t$ be the filtration
\begin{equation}
	\mathcal{F}_t=\underbrace{2^\mathcal{S}\times 2^\mathcal{S}\times \cdots2^\mathcal{S}}_\textrm{$t+1$}\times \big\{\emptyset, \mathcal{S} \big\}\times \big\{\emptyset, \mathcal{S} \big\}\times \cdots.
\end{equation}
Here by saying $f_t$ is randomly drawn, it means $f_t(\omega)=\omega_t$ and therefore  $\sigma(f_t)\in\mathcal{F}_t$.

\subsection{Lie Algebra and Lie Group}

A Lie algebra $\mathcal{L}\subset\mathscr{L}_*(\HH_N)$ is a linear subspace of $\mathscr{L}_*(\HH_N)$ which is closed under the Lie bracket operation, i.e., if $\A,\B\in\mathcal{L}$, then $[\A,\B]=\A\B-\B\A\in\mathcal{L}$.
For $\{\B_1,\dots,\B_p\}$ being a subset of $\mathcal{L}$, the Lie algebra generated by $\{\B_1,\dots,\B_p\}$, denoted by  $\mathcal{L}\{\B_1,\dots,\B_p\}$, is the smallest Lie subalgebra in $\mathscr{L}_*(\HH_N)$ containing $\{\B_1,\dots,\B_p\}$.
Given a Lie algebra $\mathcal{L}$, the associated Lie group, denoted by $\{e^\mathcal{L}\}_G$ or simply $e^\mathcal{L}$, is the one-parameter group
$\{\exp(t\A):t\in\R, \A\in\mathcal{L}\}$. Here $\exp:\mathcal{L}\rightarrow e^\mathcal{L}$ denotes the exponential map, i.e., $\exp(t\A)=e^{t\A}:=\sum_{i=0}^\infty \frac{t^i\A^i}{i!}$.

The space of skew-Hermitian operators over $\HH_N$ forms a Lie algebra, which is denoted by $\ula(N)$. The Lie group associated with $\ula(N)$ is denoted by $\ULG(N)$, which is the space of unitary operators over $\HH_N$. Let $\su(N)$ denote the Lie algebra containing all traceless skew-Hermitian operators over $\HH_N$, and $\spla(2N)$ be the Lie algebra containing $\{\X\in\su(2N):\X\mathsf{J}+\mathsf{J}\X^\top=0\}$ with $\mathsf{J}\in\mathscr{L}_*(\HH_{2N})$ whose matrix representation can be $J=\begin{pmatrix}
0 & I_N \\ -I_N & 0
\end{pmatrix}$ under certain basis.

\begin{theorem}\cite{alb03}\label{thm:psc}
	The pure state controllability and    equivalent state controllability are equivalent for the system \eqref{eq:sch-control}.  The system \eqref{eq:sch-control} is pure state controllable or equivalent state controllable if and only if $\mathcal{L}\{\A,\B_1,\dots,\B_p\}$ is isomorphic to
	\begin{align}\label{eq:isomorphic.condition}
		\begin{cases}
			\spla(N/2)~\text{or}~\su(N), & \text{$N$ is even},\\
			\su(N),                   & \text{$N$ is odd}.
		\end{cases}
	\end{align}
\end{theorem}

\section{The Quantum Network Model}\label{secmodel}

In this section, we present the quantum networks model for our study. We consider a network of qubits subject to bilinear control, which aligns with the spin-network models in the literature. We also consider a sequential measurement process where global or local qubit measurements take place periodically.

\subsection{Qubit Networks}

Qubit is the simplest quantum system whose state space is a two-dimensional Hilbert space $\HH$ ($:=\HH_2$). Let $n$ qubits indexed by $\mathrm{V}=\{1,\dots,n\}$ form a network with state space $\Hn$. The (pure) states of the qubit network are then
in the space $\QQ(2^n):=\{q\in\HH_{2^n}:|q|^2=1\}$.

Let there be a projective measurement (or an observable) for a single qubit as
$$
\M=\lambda_0\P_0+\lambda_1\P_1,
$$
where $\P_m=\braket{v_m}{v_m}$ is the projector onto the eigenspace generated by $\bra{v_m}$ with eigenvalue $\lambda_m$, $m\in\{0,1\}$.
For the $n$-qubit network, we can have either global or local measurements.

\begin{definition} (i) We term 
	$
	\M^{\otimes n}={\M\otimes \dots\otimes \M}
	$	
	as a global measurement over the $n$-qubit network.
	
	(ii) 
	Let $\mathrm{V}_*=\{i_1,\dots,i_k\}\subset\mathrm{V}$. Then 
	$$
	\M^{\mathrm{V}_*}={\mathrm{I}\otimes\dots\otimes \mathrm{I}\otimes\overbrace{\M}^{i_1\text{-th}}\otimes \mathrm{I}\otimes\cdots\otimes \mathrm{I}\otimes\overbrace{\M}^{i_k\text{-th}}\otimes \mathrm{I}\otimes\dots\otimes \mathrm{I}}
	$$
	is defined as a local measurement over $\mathrm{V}_*$.
\end{definition}

The global measurement $\M^{\otimes n}$ measures the individual qubit states of the entire network, which yields  $2^n$ possible outcomes $[\lambda_{m_1},\dots,\lambda_{m_n}],{m_j}\in\{0,1\}, j=1,\dots,n$.
Upon measuring the state $\bra{\varphi}$, the probability of getting result $[\lambda_{m_1},\dots,\lambda_{m_n}]$ is given by
$
p([\lambda_{m_1},\dots,\lambda_{m_n}])=\ket{\varphi}\P_{m_1}\otimes\cdots\otimes\P_{m_n}\bra{\varphi}.
$
Given that the outcome $[\lambda_{m_1},\dots,\lambda_{m_n}]$ occurred, the qubit network state immediately after the measurement is
$
\bra{\varphi}_\rmp=
\bra{v_{m_1}}\otimes\cdots\otimes\bra{v_{m_n}}
$.
On the other hand, the local measurement  $\M^{\mathrm{V}_*}$ measures the states of the qubits in the set  $\mathrm{V}_*$ only, which yields  $2^k$ possible outcomes $[\lambda_{m_{i_1}},\dots,\lambda_{m_{i_k}}],
i_j\in\{0,1\},j=1,\dots,k$ corresponding to the qubits $\{i_1,\dots,i_k\}$.
Upon measuring the state $\bra{\varphi}$, the probability of getting result $[\lambda_{m_{i_1}},\dots,\lambda_{m_{i_k}}]$ is 
\begin{align*}
	&p([\lambda_{m_{i_1}},\dots,\lambda_{m_{i_k}}])=\ket{\varphi}\mathrm{I}\otimes\dots\otimes \mathrm{I}\otimes{\P_{m_{i_1}}}\\
	&\otimes \mathrm{I}\otimes\cdots\otimes \mathrm{I}\otimes{\P_{m_{i_k}}}\otimes \mathrm{I}\otimes\dots\otimes \mathrm{I}\bra{\varphi},\quad 
\end{align*}
where $m_{i_j}\in\{0,1\},j=1,\dots,k$.
Since the local measurement   reveals no information about the nodes in $\mathrm{V}\setminus\mathrm{V}_*$, we term the qubits in $\mathrm{V}_*$ as the measured qubits, and those in $\mathrm{V}\setminus\mathrm{V}_*$ as the dark qubits. For the ease of presentation and without loss of generality, we assume $\mathrm{V}_*=\{1,\dots,k\}$ throughout the remainder of the paper.

\subsection{Hybrid Qubit Network Dynamics}

Consider the continuous time horizon represented by $s\in[0,\infty)$. Let $\bra{q(s)}$ denote the qubit network state at time $s$. Let the evolution of $\bra{q(s)}$ be defined by a Schr\"odinger equation with controlled Hamiltonians in the form of \eqref{eq:sch-control}, and the network state be measured globally or locally from $s=0$ periodically with a period $T$.
To be precise, $\bra{q(s)}$ satisfies the following hybrid dynamical equations
\begin{align}
	\label{eq:hybrid.network}
	&\bra{\dot q(s)} = \left(\A+\sum_{\ell=1}^p u_\ell(s)\B_\ell\right) \bra{q(s)},\quad s\in[tT,(t+1)T), \\
	\label{eq:right.state}
	&\bra{q((t+1)T)}=\bra{q((t+1)T)^-}_\rmp,
\end{align}
for $t=0,1,2,\cdots$, where $\bra{q((t+1)T)^-}$ represents the quantum network state right before $(t+1)T$ along \eqref{eq:hybrid.network} starting from $\bra{q(tT)}$, and $\bra{q((t+1)T)^-}_\rmp$ is the post-measurement state of the network when a measurement is performed at time $s=(t+1)T$.
For the ease of presentation, we define quantum states
$$
\begin{aligned}
&\bra{\psi(t)}=\bra{q((tT)^-)},\\
&\bra{\psi(t)}_\rmp=\bra{q(tT)}
\end{aligned}
$$
for the pre- and post-measurement network states at the $(t+1)$-th measurement.

In particular, the control signals $u_\ell(s)$, $\ell=1,\dots,p$ will have feedforward or feedback forms. 

\begin{definition}
	(i) The control signals $u_\ell(s)$, $\ell=1,\dots,p$ are feedforward if their values are determined deterministically at $t=0^-$ for the entire time horizon $s\geq 0$. 
	
	(ii) The control signals $u_\ell(s)$, $\ell=1,\dots,p$ are feedback if each $u_\ell(s)$ for $s\in[tT,(t+1)T)$ depends on the post-measurement state $\bra{\psi(t^\prime)}_\rmp$, $t^\prime=0,1,\dots,t$.
\end{definition}

 \bluetext{
	\subsection{Problems of Interest}
	The evolution of the quantum system (\ref{eq:hybrid.network})--(\ref{eq:right.state}) defines a quantum hybrid with state resets, analogous to the study of classical hybrid systems with state jumps~\cite{sicon2014}.  We note that such state evolution represents physical systems that exist in the real world, where sequential measurements are performed for quantum dynamical systems \cite{QubitFeedback2014}.  The mixture of the continuous-time dynamics and the random state resets leads to intrinsic questions related to the relationship between the quantum state and the measurement outcome evolutions. Furthermore, how  the continuous bilinear control  (\ref{eq:hybrid.network}) will be affected by the sequential measurements is also an interesting point for investigation. In this paper, we focus on the following questions:
	\begin{itemize}
		\item[Q1:] How can we characterize the dynamics of the measurement outcomes from the quantum networks with feedforward control?
		
		\item[Q2:] How the sequential measurements with feedback control will influence the controllability properties of the classical bilinear model \eqref{eq:hybrid.network}?    
	\end{itemize}
}

\section{Boolean Dynamics from Quantum Measurements}\label{secboolean}

In this section, we focus our attention on the induced Boolean dynamics from the sequential measurements of the qubit networks. We impose the following assumption.

\begin{assumption}\label{ass:1}
	The $u_\ell(s)$, $\ell=1,\dots,p$ are feedforward signals. Consequently, there exist a sequence of deterministic $\U_t$, $t=0,1,2,\dots$ such that $\bra{\psi(t+1)}=\U_t\bra{\psi(t)}_\rmp$.
\end{assumption}

\subsection{Induced Probabilistic Boolean Networks}

Under the global measurement $\M^{\otimes n}$, 
we can use the Boolean variable  $x_i(t)\in\{0,1\}$    to represent the measurement outcome at qubit $i$ for step $t$, where  $x_i(t)=0$ corresponds to $\lambda_0$ and $x_i(t)=1$ corresponds to $\lambda_1$. We can further define the $n$-dimensional random Boolean vector
$$
\mathbf{x}(t)=[x_1(t),\cdots,x_n(t)]\in \{0,1\}^{n}
$$
as the outcome of measuring $\bra{\psi(t)}$ 
under $\M^{\otimes n}$ at step $t$.  The recursion of $\bra{\psi(t)}_\rmp$ generates the corresponding recursion of $\mathbf{x}(t)$ for $t=0,1,2,\dots$, resulting in an induced probabilistic Boolean network (PBN).
Similarly, subject to local measurement, we can define 
$
\mathbf{x}_{\mathrm{k}}(t)=[x_{1}(t),\cdots,x_{k}(t)]\in \{0,1\}^k
$
as the outcome of measuring $\bra{\psi(t)}$ by $\M^{\mathrm{V}_\ast}$, where $x_i(t)\in\{0,1\}$ continues to represent the measurement outcome at qubit $i$. 

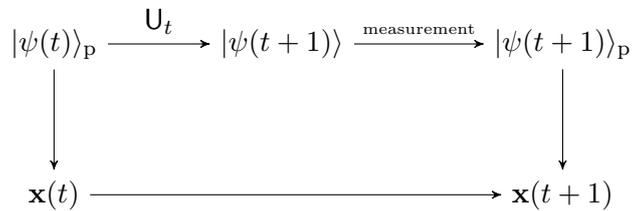
\begin{figure}[!htb]
	\centering
	\begin{tikzpicture}[->,>=stealth',auto,node distance=3cm]
	\tikzstyle{state}=[semithick,minimum size=6mm]
	\node[state] (q2) {$\bra{\psi(t)}_\rmp$};
	\node[state,right of=q2] (q3) {$\bra{\psi(t+1)}$};
	\node[right of=q3,xshift=1.8em] (q4) {$\bra{\psi(t+1)}_\rmp$};
	\node[state,below of=q2,yshift=1cm] (x1) {$\mathbf{x}(t)$};
	\node[state,below of=q4,yshift=1cm] (x2) {$\mathbf{x}(t+1)$};
	\path (q2) edge node[above]{$\U_t$}                           (q3);
	\path (q3) edge node[above]{\tiny measurement} (q4);
	\path[->] (q2) edge (x1);
	\path[->] (q4) edge (x2);
	\path (x1) edge node[above]{} (x2);
	\end{tikzpicture}
	\caption{Induced Boolean network dynamics.}\label{fig:induced.boolean.map}
\end{figure}

We are interested in the interplay between the underlying quantum state evolution and the induced probabilistic Boolean network dynamics.

\subsection{Global Measurement: Markovian PBN}

\subsubsection{Transition Characterizations}

We first analyze the behaviors of the induced probabilistic Boolean network dynamics under global qubit network measurements. Let $\delta_N^i$ be the $i$-th column of identify matrix $I_N$. 
Denote  $\Delta_N=\{\delta_N^i|i=1,\dots,N\}$, and particularly  $\Delta:=\Delta_2$ for simplicity.
Identify $\{0,1\}\simeq\Delta$ under which $0\sim\delta_2^1$ and $1\sim\delta_2^2$.
Let $\mathbf{x}=[x_1,\dots,x_n]\in \{0,1\}^n$ be associated with  
\begin{align}\label{eq:composite.boolean.map}
	\mathbf{x}^\sharp:=\delta_{2}^{x_1+1}\otimes\cdots\otimes\delta_{2}^{x_n+1}=\delta_{2^n}^{\sum_{i=1}^n x_i2^{n-i}+1},
\end{align}
where $\otimes$ represents the Kronecker product. In this way, we have identified $\{0,1\}^n\simeq\Delta_{2^n}$.
For the ease of presentation, we also denote $\lfloor\mathbf{x}\rfloor:=\sum_{i=1}^n x_i2^{n-i}+1$, and consider $\mathbf{x}$, $\lfloor\mathbf{x}\rfloor$, and  $\mathbf{x}^\sharp=\delta_{2^n}^{\lfloor\mathbf{x}\rfloor}$ interchangeable without further mentioning. Recall $\mathcal{S}$ as the set containing all $(2^n)^{2^n}$ Boolean mappings from $\{0,1\}^n$ to $\{0,1\}^n$. Each element in $\mathcal{S}$ is indexed by
$f_{[\alpha_1,\dots,\alpha_{2^n}]}\in\mathcal{S}$ with  $\alpha_i=1,\dots,2^n,i=1,\dots,2^n$,
where
\begin{align}\label{eq:ft1}
	f_{[\alpha_1,\dots,\alpha_{2^n}]}(s_{i})=s_{\alpha_i},\quad s_i\in \{0,1\}^n,~i=1,\dots,2^n.
\end{align}
In this way,  the matrix $f_{[\alpha_1,\dots,\alpha_{2^n}]}=\left[\delta_{2^n}^{\alpha_1},\dots,\delta_{2^n}^{\alpha_{2^n}}\right]$ serves as a representation of $f_{[\alpha_1,\dots,\alpha_{2^n}]}$ since
\begin{align}\label{eq:Ft}
	f_{[\alpha_1,\dots,\alpha_{2^n}]}\delta_{2^n}^{i}=\delta_{2^n}^{\alpha_i},\quad i=1,\dots,2^n. 
\end{align}

Recall the observable $\M=\lambda_0\P_0+\lambda_1\P_1$ for one qubit. We choose $\{\bra{0},\bra{1}\}$ as the standard orthonormal basis of $\HH$, and denote $\Q_0=\braket{0}{0}$, $\Q_1=\braket{1}{1}$. Then there exists a unitary operator $\uu=\braket{v_0}{0}+\braket{v_1}{1}\in\mathfrak{L}_*(\HH)$, whose representation under the chosen basis $\{\bra{0},\bra{1}\}$ is $u\in\mathbb{C}^{2\times 2}$ which is a unitary matrix, such that $\P_0=\uu\Q_0\uu^\dag$ and $\P_1=\uu\Q_1\uu^\dag$. 

Let $\{\bra{0},\bra{1}\}^{\otimes n}$ be the standard computational basis of the $n$-qubit network. We denote for $i=1,\dots,2^n$ that
\begin{align}\label{basis}
	\bra{b_i}=\bra{b_{i_1}\cdots b_{i_n}}
\end{align}
where $\bra{b_{i_1}\cdots b_{i_n}}\in \{\bra{0},\bra{1}\}^{\otimes n}$ with   $b_{i_j}\in\{0,1\},~j=1,\dots,n$. Now we can sort the elements of $\{\bra{0},\bra{1}\}^{\otimes n}$ by the value of $\lfloor b_i\rfloor$ in an ascending order.
Let $\U_t$ have the representation  $U_t\in\mathbb{C}^{2^n\times 2^n}$ under such an ordered basis. Note that $\mathsf{u}\otimes\dots\otimes \mathsf{u}$ has its matrix representation as $u\otimes\dots\otimes u$ under the same sorted basis.  Define
\begin{align}
	U_t^\M=({u\otimes\dots\otimes u})^\dag U_t({u\otimes\dots\otimes u}).
\end{align}
For the induced Boolean series $\{\mathbf{x}(t)\}_{t=0}^\infty $, the following result holds, whose proof is omitted as it is a direct verification of quantum measurement postulate.   

\begin{proposition}\label{prop0}
	Let Assumption \ref{ass:1} hold.
	With global measurement,  the $\{\mathbf{x}(t)\}_{t=0}^\infty $ form a  Markov chain over the state space $\{0,1\}^n$, whose state transition matrix $\mathbf{P}_t$ at time $t$ is given by
	$$
	[\mathbf{P}_t]_{i,j}=\mathbb{P}\Big(\mathbf{x}(t+1)\Big|\mathbf{x}(t)\Big)=\Big| [U_t^\M]_{j,i}\Big|^2,\quad 
	$$
	for $i=\lfloor\mathbf{x}(t)\rfloor,j=\lfloor\mathbf{x}(t+1)\rfloor\in\{1,2,\dots,2^n\}$, 
	where $ [\cdot]_{i,j}$ stands for the $(i,j)$-th entry of a matrix.
	In fact, there holds
	$
	\mathbf{P}_t= (U_t^\M)^\dag\circ (U_t^\M)^\top,
	$
	where $\circ$ stands for the Hadamard product.
\end{proposition}

The following theorem establishes an algebraic representation of the recursion for $\{\mathbf{x}(t)\}_{t=0}^\infty $.  
\begin{theorem}\label{thm:1}
	Let Assumption \ref{ass:1} hold.
	The recursion of $\{\mathbf{x}(t)\}_{t=0}^\infty $ can be represented as  a random linear mapping 
	\begin{align}\label{eq:linear.mapping}
		\mathbf{x}^\sharp(t+1)= F_t  \mathbf{x}^\sharp(t), 
	\end{align}
	where $\left<F_t\right>$  is a series of  independent random matrices in $\mathbb{R}^{2^n\times 2^n}$. Moreover, the distribution of $F_t$ is described by
	$$
	\mathbb{P}(F_t=f_{[\alpha_1,\dots,\alpha_{2^n}]})=\prod\limits_{i=1}^{2^n}\Big|[U_t^\M]_{\alpha_i,i}\Big|^2.
	$$
\end{theorem}

 \bluetext{
	The proofs of Proposition~\ref{prop0} and Theorem~\ref{thm:1} are deferred to the Appendix.
	
	\begin{remark}
		Although Theorem~\ref{thm:1} provides a way of explicitly representing the evolution of the measurement outcomes, the inherent computational complexity does not get reduced. The dimension of $\mathbf{x}^\sharp(t)$ grows exponentially as the number of qubits grows. However, the state transition $F_t$ is in general a sparse matrix, which might lead to potential computational reduction in the establishment on usage of \eqref{eq:linear.mapping}.
	\end{remark}
	
	\begin{remark}
		Note that Proposition \ref{prop0} and Theorem \ref{thm:1} hold for general quantum states and unitary evolution $\mathsf{U}_t$. Let $\mathsf{M}$ be taken as the standard computational basis. Then from the identity $\mathbf{P}_t= (U_t)^\dag\circ (U_t)^\top$, the structure of $\mathsf{U}_t$ is fully inherited by $\mathbf{P}_t$. As a result, if $\mathsf{U}_t$ is an entangling unitary operator, the same entangling structure will be preserved by the state-transition matrix $\mathbf{P}_t$. In fact, the correlations between the $x_i(t)$ arise from $\mathbf{P}_t$, in contrast to the correlation of the qubit states induced by $\mathsf{U}_t$.  
	\end{remark}
}

\subsubsection{Quantum Realization of Classical PBN}

From Theorem \ref{thm:1}, one can see that the $n$-qubit network under global sequential measurement $\M^{\otimes n}$ always induces a Markovian probabilistic Boolean network. When $\U_t$ is time invariant,  $\{\mathbf{x}(t)\}_{t=0}^\infty$ is a homogeneous chain. A natural question lies in  whether any classic probabilistic Boolean network with a homogeneous transition could be realized by the qubit networks under investigation. 
This question is related to the unistochastic matrix theory. 
A matrix $W\in\R^{N\times N}$ is doubly stochastic if it is a square matrix of nonnegative real numbers, each of whose rows and columns sums to $1$, i.e., $\sum_i [W]_{i,j}=\sum_j [W]_{i,j}=1$. 
A doubly stochastic matrix $T$ is unistochastic if its entries are the squares of the absolute values of the entries from certain unitary matrix, i.e., there exists a unitary matrix $U$ such that $[W]_{i,j}=|[U]_{i,j}|^2$ for $i,j=1,\dots,N$. It is still an open problem to tell whether a given doubly stochastic matrix is unistochastic or not~\cite{JMP-2009-Dunkl}.

Note that instead of using the global measurement $\M^{\otimes n}$, we may choose   another global measurement as $\M_1\otimes\dots\otimes\M_n$, i.e., the observable of qubit $i$ is $\M_i=\lambda_{i_0}\braket{v_{i_0}}{v_{i_0}}+\lambda_{i_1}\braket{v_{i_1}}{v_{i_1}}$, then assume the matrix representation of $\uu_i=\braket{v_{i_0}}{0}+\braket{v_{i_1}}{1}$ is $u_i$ for qubit $i$ under the basis $\{\bra{0},\bra{1}\}$. Then we have
$U^\M=(u_1\otimes\cdots\otimes u_n)^\dag U (u_1\otimes\cdots\otimes u_n)$, which is still a unitary matrix. As a result, using a more general measurement $\M_1\otimes\dots\otimes\M_n$ does not reduce the difficulty of the quantum realization problem.

Alternatively, we can try to solve the quantum realization problem approximately.
Given a column stochastic matrix $W\in\R^{N\times N}$, we define
\begin{align*}
	\begin{array}{cl}
		\text{minimize}&\sum\limits_{i,j=1,\dots,N}\Big||[U]_{i,j}|^2-[W]_{i,j}\Big|^2\\
		\text{subject to}&UU^\dag=I,~~U\in\C^{N\times N},
	\end{array}
\end{align*}
which is a polynomial optimization problem. 

 \bluetext{
	In general, this optimization problem may lead to multiple solutions, implying potential ambiguity in identifying the unitary operator  $\mathsf{U}$ from the state-transition probability matrix of the induced Markov chain. However, whenever such an optimization problem yields exact solutions, or a solution with a sufficiently small gap compared to exact solutions,  our quantum network with sequential measurements becomes a potential resource for the realization of the given Markov chain. For a Markov chain with $L$ states, it suffices to use $\log L$ qubits for the quantum network realization, where the quantum measurements become the intrinsic resource of the randomness.  }

\subsubsection{Examples}

We consider a two-qubit network. Let an observable be given for one qubit along standard computational basis $\{\bra{0},\bra{1}\}$ as $\M=\lambda_0\braket{0}{0}+\lambda_1\braket{1}{1}$. The resulting  global network measurement is $\M\otimes \M$. Then the set of possible outcomes is $\{0,1\}^2$. The random Boolean mapping $F_t:\mathcal{S}\rightarrow \mathcal{S}$ has $4^4=256$ possible realizations.

\begin{example}\normalfont
	Let the unitary operator  acting on the two-qubit network  be
	$$
	\U_t\equiv\U_1=(\braket{0}{1}+\braket{1}{0})\otimes(-\imath\braket{0}{1}+\imath\braket{1}{0}).
	$$
	The state transition map of the homogeneous Markov chain induced by $\U_1$ and $\M$ is shown in Fig.~\ref{fig:ex1}, and $F_t$ has only one realization.
	\begin{figure}[!htb]
		\centering
		\begin{tikzpicture}[->,>=stealth',auto,node distance=3cm]
		\tikzstyle{state}=[draw=black,circle,thick,minimum size=6mm]
		\node[state] (00)             {$\bra{00}$};
		\node[state,right of=00] (01) {$\bra{01}$};
		\node[state,below of=00] (10) {$\bra{10}$};
		\node[state,right of=10] (11) {$\bra{11}$};
		\path (00.305) edge [bend right] node[right,near end,xshift=2mm]{$1$} (11.155);
		\path (11.135) edge [bend right] node[right,near end,xshift=-7mm]{$1$} (00.325);
		\path (01.215) edge [bend right] node[right,near end,xshift=-2mm,yshift=-3mm]{$1$} (10.55);
		\path (10.35) edge [bend right] node[right,near end,xshift=-3mm,yshift=3mm]{$1$} (01.235);
		\end{tikzpicture}
		\caption{State transition map of the Markov chain induced by $\U_1$ and $\M$.}\label{fig:ex1}
	\end{figure}
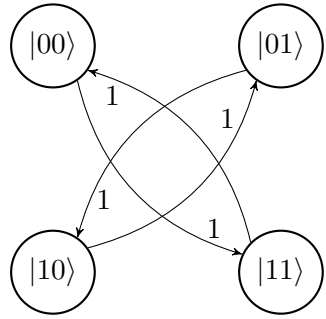
\end{example}

\begin{example}
	\normalfont
	Let the unitary operator    be alternatively given as
	$$
	\U_t\equiv\U_2=(\braket{0}{1}+\braket{1}{0})\otimes\frac{\braket{0}{0}+\braket{0}{1}-\braket{1}{0}+\braket{1}{1}}{\sqrt{2}}. 
	$$
	The state transition map of the homogeneous Markov chain induced by $\U_2$ and $\M$ is shown in Fig.~\ref{fig:ex2}. Moreover, $F_t$ has $16$ realizations each of which happens with equal probability $1/16$.
	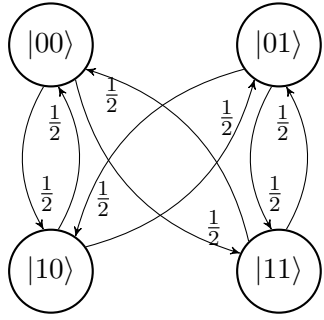
\begin{figure}[!htb]
		\centering
		\begin{tikzpicture}[->,>=stealth',auto,node distance=3cm]
		\tikzstyle{state}=[draw=black,circle,thick,minimum size=6mm]
		\node[state] (00)             {$\bra{00}$};
		\node[state,right of=00] (01) {$\bra{01}$};
		\node[state,below of=00] (10) {$\bra{10}$};
		\node[state,right of=10] (11) {$\bra{11}$};
		\path (00.260) edge [bend right] node[right,near end]{$\frac{1}{2}$} (10.100);
		\path (10.80) edge [bend right] node[left,near end]{$\frac{1}{2}$} (00.280);
		\path (01.260) edge [bend right] node[right,near end]{$\frac{1}{2}$} (11.100);
		\path (11.80) edge [bend right] node[left,near end]{$\frac{1}{2}$} (01.280);
		\path (00.305) edge [bend right] node[right,near end,xshift=2mm]{$\frac{1}{2}$} (11.155);
		\path (11.135) edge [bend right] node[right,near end,xshift=-7mm]{$\frac{1}{2}$} (00.325);
		\path (01.215) edge [bend right] node[right,near end,xshift=-2mm,yshift=-3mm]{$\frac{1}{2}$} (10.55);
		\path (10.35) edge [bend right] node[right,near end,xshift=-3mm,yshift=3mm]{$\frac{1}{2}$} (01.235);
		\end{tikzpicture}
		\caption{State transition map of the Markov chain induced by $\U_2$ and $\M$.}\label{fig:ex2}
	\end{figure}
	
\end{example}

 \bluetext{
	\begin{example}\normalfont
		Let $\oH=\frac{\pi}{3}(\braket{0}{1}+\braket{1}{0})\otimes(\braket{0}{1}+\braket{1}{0})+\frac{\pi}{6}(-\imath\braket{0}{1}+\imath\braket{1}{0})\otimes(-\imath\braket{0}{1}+\imath\braket{1}{0})$. 
		Then
		$$
		\begin{aligned}
		e^{-\imath\oH}=&\frac{\sqrt{3}}{2}\braket{00}{00}-\imath\frac{1}{2}\braket{00}{11}-\imath\braket{01}{10}\\
		&-\imath\braket{10}{01}-\imath\frac{1}{2}\braket{11}{00}+\frac{\sqrt{3}}{2}\braket{11}{11}.
		\end{aligned}
		$$
		is an entangling unitary operator (e.g., \citeasnoun{cohen2011}).  Let $\U_t\equiv\U_3=e^{-\imath\oH}$ for all $t=0,1,2,\dots$.
		The state transition map of the Markov chain induced by $\U_3$ and $\M$ is shown in Fig.~\ref{fig:ex3}. Also, the state transition maps for each qubit when the two-qubit network starts from the state $\bra{00}$ are shown in Fig.~\ref{fig:ex3-2}. 
		
		As we can see, starting from the product state $\bra{00}$ and after the operation of $\U_3$, the measurement outcomes $\mathbf{x}_1(t)$ and $\mathbf{x}_2(t)$ become statistically correlated. The entangling relationship generated by $\U_t$ is then reflected in the state transition of the induced Boolean dynamics.
		
		\begin{figure}[!htb]
			\centering
			\begin{tikzpicture}[->,>=stealth',auto,node distance=3cm]
			\tikzstyle{state}=[draw=black,circle,thick,minimum size=6mm]
			\node[state] (00)             {$\bra{00}$};
			\node[state,right of=00] (01) {$\bra{01}$};
			\node[state,below of=00] (10) {$\bra{10}$};
			\node[state,right of=10] (11) {$\bra{11}$};
			\path (00.305) edge [bend right] node[right,near end,xshift=2mm]{$\frac{1}{4}$} (11.155);
			\path (11.135) edge [bend right] node[right,near end,xshift=-7mm]{$\frac{1}{4}$} (00.325);
			\path (01.215) edge [bend right] node[right,near end,xshift=-2mm,yshift=-3mm]{$1$} (10.55);
			\path (10.35) edge [bend right] node[right,near end,xshift=-3mm,yshift=3mm]{$1$} (01.235);
			\path (00) edge [loop left] node[left]{$\frac{3}{4}$} (00);
			\path (11) edge [loop right] node[right]{$\frac{3}{4}$} (11);
			\end{tikzpicture}
			\caption{State transition map of the Markov chain induced by $\U_3$ and $\M$.}\label{fig:ex3}
		\end{figure}
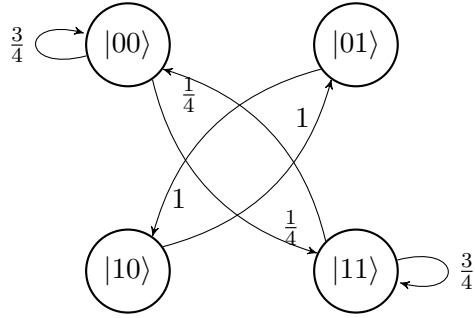
		
		\begin{figure}[!htb]
			\centering
			\begin{tikzpicture}[->,>=stealth',auto,node distance=2.5cm]
			\tikzstyle{state}=[draw=black,circle,thick,minimum size=6mm]
			\node[state] (a0)             {$\bra{0}_1$};
			\node[state,right of=a0] (a1) {$\bra{1}_1$};
			\path (a0) edge [loop above] node[above]{$\frac{3}{4}$} (a0);
			\path (a1) edge [loop above] node[above]{$\frac{3}{4}$} (a1);
			\path (a0.5) edge [bend left] node[above]{$\frac{1}{4}$} (a1.175);
			\path (a1.185) edge [bend left] node[below]{$\frac{1}{4}$} (a0.355);
			\node [below of=a0,xshift=1.25cm,yshift=1cm] {Qubit 1};
			\begin{scope}[xshift=4.5cm]
			\node[state] (b0)             {$\bra{0}_2$};
			\node[state,right of=b0] (b1) {$\bra{1}_2$};
			\path (b0) edge [loop above] node[above]{$\frac{3}{4}$} (b0);
			\path (b1) edge [loop above] node[above]{$\frac{3}{4}$} (b1);
			\path (b0.5) edge [bend left] node[above]{$\frac{1}{4}$} (b1.175);
			\path (b1.185) edge [bend left] node[below]{$\frac{1}{4}$} (b0.355);
			\node [below of=b0,xshift=1.25cm,yshift=1cm] {Qubit 2};
			\end{scope}
			\end{tikzpicture}
			\caption{State transition maps of individual qubits starting from $\bra{00}$ in the Markov chain induced by $\U_3$ and $\M$.}\label{fig:ex3-2}
		\end{figure}
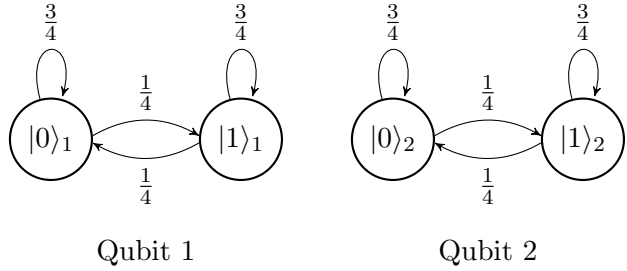
	\end{example}
}

\begin{example}\normalfont
	Consider the following doubly stochastic matrix in $\R^{4\times 4}$
	$$
	W=\begin{pmatrix}
	\frac{1}{12} & \frac{1}{6} & \frac{1}{4} & \frac{1}{2} \\
	\frac{1}{6} & \frac{1}{12} & \frac{1}{2} & \frac{1}{4} \\
	\frac{1}{4} & \frac{1}{2} & \frac{1}{12} & \frac{1}{6} \\
	\frac{1}{2} & \frac{1}{4} & \frac{1}{6} & \frac{1}{12} \\
	\end{pmatrix}.
	$$
	Then we can find the following unitary matrix
	$$
	U=\begin{pmatrix}
	\frac{1}{2\sqrt{3}}& \frac{1}{\sqrt{6}} & \frac{1}{2} & \frac{\sqrt{2}}{2}\\
	-\frac{1}{\sqrt{6}}i& \frac{1}{2\sqrt{3}}i & -\frac{\sqrt{2}}{2}i & \frac{1}{2}i\\
	-\frac{1}{4}-\frac{\sqrt{3}}{4}i& -\frac{\sqrt{2}}{4}-\frac{\sqrt{6}}{4}i& \frac{1}{4\sqrt{3}}+\frac{1}{4}i & \frac{1}{2\sqrt{6}}+\frac{\sqrt{2}}{4}i\\
	-\frac{\sqrt{6}}{4}+\frac{\sqrt{2}}{4}i & \frac{\sqrt{3}}{4}-\frac{1}{4}i & \frac{\sqrt{2}}{4}-\frac{1}{2\sqrt{6}}i & -\frac{1}{4}+\frac{1}{4\sqrt{3}}i \\
	\end{pmatrix},
	$$
	such that $U^\dag\circ U^\top=W$.
	
	Let a Markov chain over a four-state space $\{s_1,s_2,s_3,s_4\}$ with state transition matrix $W$ evolve from initial distribution $p_0=\left(\frac{1}{2},\frac{1}{6},\frac{1}{12},\frac{1}{4}\right)^\top$. 
	Let $\M^{\otimes 2}$ be the measurement of a qubit network.
	We encode $s_1\simeq \bra{00}$, $s_2\simeq \bra{01}$, $s_3\simeq \bra{10}$, $s_4\simeq \bra{11}$.
	Let the qubit network start from 
	$$
	\bra{\psi(0)}=\frac{1}{\sqrt{2}}\bra{00}+\frac{1}{\sqrt{6}}\bra{01}+\frac{1}{2\sqrt{3}}\bra{10}+\frac{1}{2}\bra{11}.
	$$
	 \bluetext{We numerically simulate  the dynamics of $\mathbf{x}(t)$ for $10^4$ rounds and therefore obtain $10^4$ independent sample paths  of  $\mathbf{x}(t)$ with the same initial condition.} Then we plot the trajectory of 
	$$
	\begin{aligned}
	\hat{p}(t)=&\left(\hat{p}_1(t),\hat{p}_2(t),\hat{p}_3(t),\hat{p}_4(t)\right)^\top\\
	:=&\Big(\mathbb{P}\left(\mathbf{x}(t)=00\right), \mathbb{P}\left(\mathbf{x}(t)=01\right), \\
	&\quad \quad \mathbb{P}\left(\mathbf{x}(t)=10\right), \mathbb{P}\left(\mathbf{x}(t)=11\right)\Big)^\top
	\end{aligned}
	$$ from the experimental data as shown in Fig.~\ref{fig:example.sim01}.  \bluetext{Here $\hat{p}_i(t)=\frac{\#\{\lfloor\mathbf{x}(t)\rfloor=i\}}{10^4}$, as an unbiased estimate of $p_i(t)$.} We can also define $$p(t)=(p_1(t),p_2(t),p_3(t),p_4(t))^\top=W^tp_0,$$ which trajectory is displayed in Fig.~\ref{fig:example.sim02}. Since it is homogeneous Markov chain, which will converge to a steady distribution, one can obtain that $\lim_{t\rightarrow\infty} p(t)=[\frac{1}{4},\frac{1}{4},\frac{1}{4},\frac{1}{4}]^\top$. From these two figures, one can easily see that $\hat{p}(t)$ is an excellent estimate of $p(t)$.
	
	\begin{figure}[!htb]
		\centering
		\includegraphics[width=.5\textwidth]{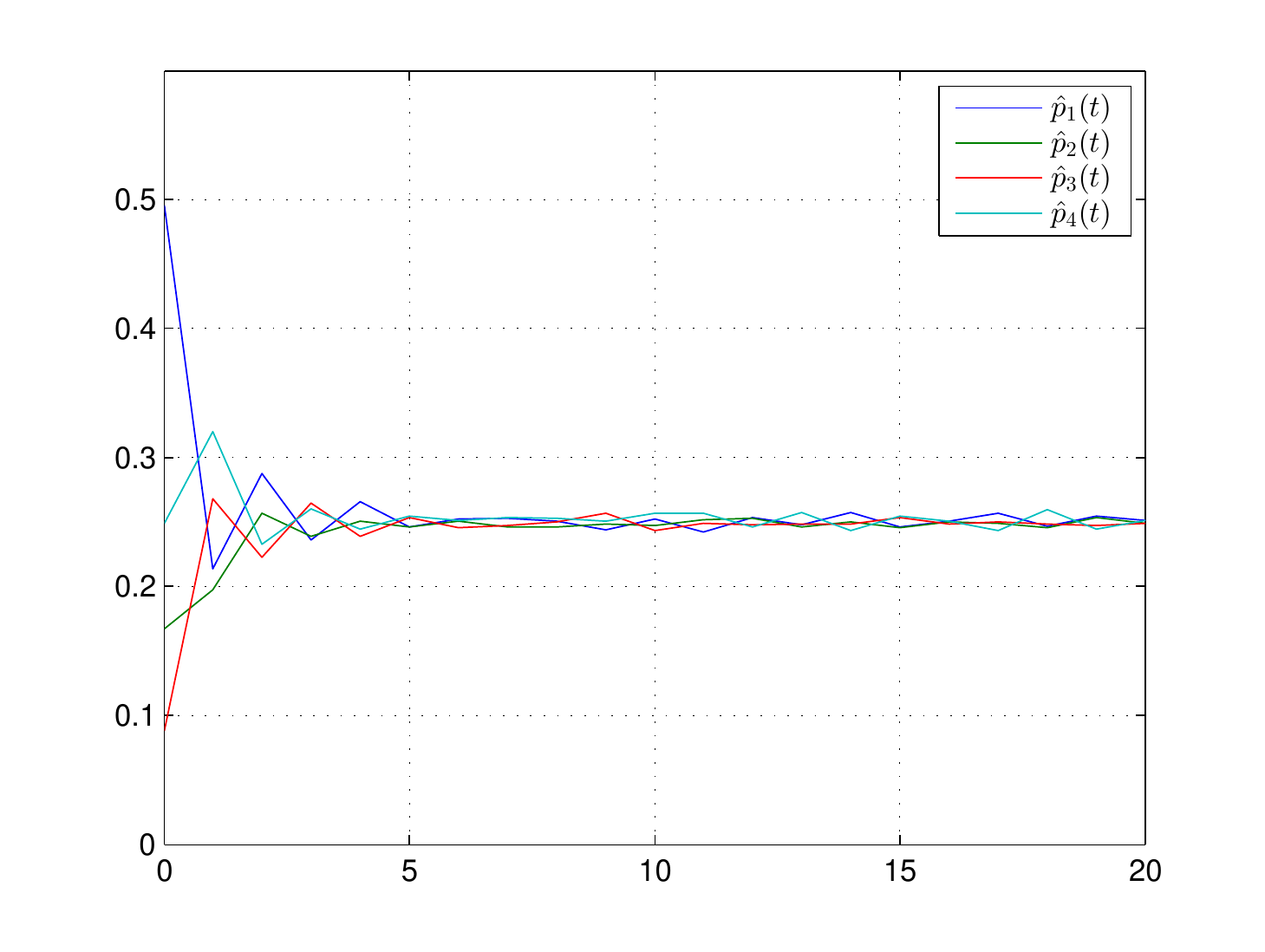}
		\caption{The trajectory of $\hat{p}(t)$ starting from the state $\bra{\psi(0)}$.}\label{fig:example.sim01}
	\end{figure}
	
	\begin{figure}[!htb]
		\centering
		\includegraphics[width=.5\textwidth]{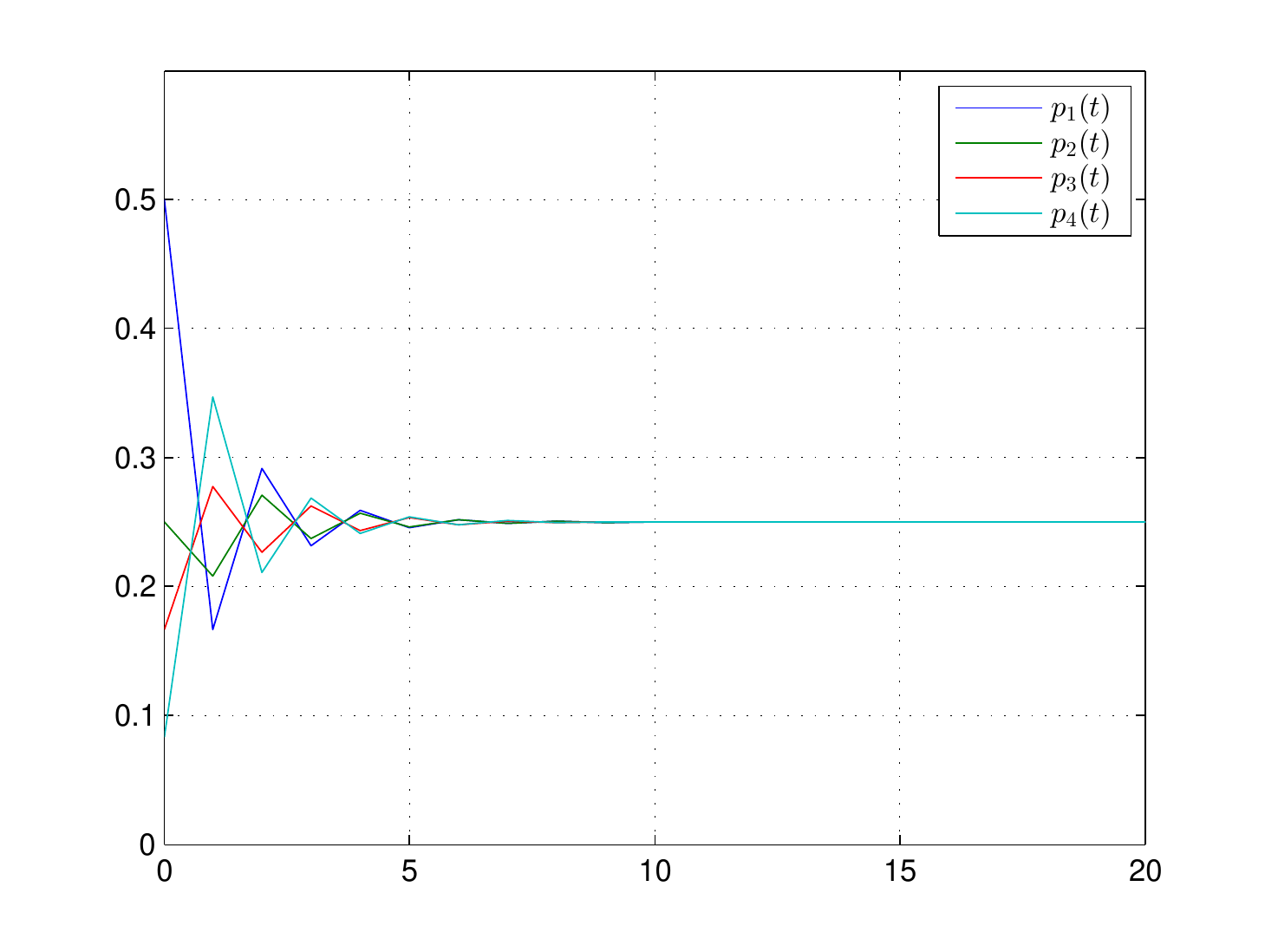}
		\caption{The trajectory of $p(t)$ starting from $p_0$.}\label{fig:example.sim02}
	\end{figure}
	
\end{example}

\subsection{Local Measurement: Non-Markovian PBN}

We now turn to the local measurement case, where at time $t$, $\M^{\mathrm{V}_*}=\M^{\otimes k}\otimes \mathrm{I}^{\otimes(n-k)}$ is performed over $\bra{\psi(t)}$ and produces outcome
$
\mathbf{x}_{\mathrm{k}}(t)=[x_1(t),\dots,x_k(t)].
$
 \bluetext{
	The operators $\U_t$ and $\M$ collectively determine  the dynamics of the quantum states and the resulting Boolean states, while any two different measurement bases $\M$ are only subject to a coordinate change. } Therefore, without loss of generality,  we assume  that
$\M=\lambda_0\P_0+\lambda_1\P_1=\lambda_0\braket{0}{0}+\lambda_1\braket{1}{1}$.

Given $\mathbf{x}_{\mathrm{k}}(t)$, the post-measurement state $\bra{\psi(t)}_\rmp$ depends on $\mathbf{x}_{\mathrm{k}}(0),\dots,\mathbf{x}_{\mathrm{k}}(t-1)$ due to the local measurement effect as $\mathbf{x}_{\mathrm{k}}(t)$ alone is not enough to determine $\bra{\psi(t)}$. Therefore $\{\mathbf{x}_{\mathrm{k}}(t)\}_{t=0}^\infty$ is no longer Markovian.
Let $\mathrm{r}:~\mathpzc{x}_{\mathrm{k}}(0),\dots,\mathpzc{x}_{\mathrm{k}}(t)$ be a path of measurement realization. Define
$$
\begin{aligned}
\PP_\mathrm{r}(0)&:=\mathbb{P}(\mathpzc{x}_{\mathrm{k}}(0))\\
\PP_\mathrm{r}(1)&:=\mathbb{P}(\mathpzc{x}_{\mathrm{k}}(1)|\mathpzc{x}_{\mathrm{k}}(0))\\
& ~~\vdots~~\\
\PP_\mathrm{r}(t+1)&:=\mathbb{P}(\mathpzc{x}_{\mathrm{k}}(t+1)|\mathpzc{x}_{\mathrm{k}}(t),\dots,\mathpzc{x}_{\mathrm{k}}(0)).
\end{aligned}
$$

We aim to provide a recursive way of calculating the above transition probabilities. 
Recall from (\ref{basis}) that   $\{\bra{0},\bra{1}\}^{\otimes n}=\{\bra{b_i}, i=1,\dots,2^n\}$ is a sorted  basis for $\HH^{\otimes n}$. 
Let $$
\bra{\psi(0)}=\sum_{i=1}^{2^n}a_i\bra{b_i}
$$ with $\sum_{i=1}^{2^n}|a_i|^2=1$ be the state of the quantum network at time $t=0$. Let $U_t$ be the matrix representation of $\U_t$ under the chosen basis for $t=0,1,2,\cdots$. Let $P_0,P_1$ be   defined in (\ref{eq100}) as the matrix representations of $\P_0,\P_1$ under the standard computational basis, respectively. 
Recall $\lfloor{\mathpzc{x}}_{\mathrm{k}}(t)\rfloor:=\sum_{i=1}^k x_i(t)2^{k-i}+1$, and $\mathpzc{x}^\sharp_{\mathrm{k}}(t):=\delta_{2^k}^{\lfloor{\mathpzc{x}}_{\mathrm{k}}(t)\rfloor}$. Then we have the following theorem.

\begin{theorem}\label{thm:local.measurement}
	Let Assumption \ref{ass:1} hold and $\M=\lambda_0\braket{0}{0}+\lambda_1\braket{1}{1}$.
	Let $\mathrm{r}:\mathpzc{x}_{\mathrm{k}}(0),\dots$, $\mathpzc{x}_{\mathrm{k}}(t)$ be a realization of the random measurement outcomes. Then there exist $\beta^\mathrm{r}(t)\in\C^{2^{n-k}}$ with $\beta^\mathrm{r}(t)=[\beta_1^\mathrm{r}(t),\dots,\beta_{2^{n-k}}^\mathrm{r}(t)]^\top$ for $t=0,1,2,\dots$, such that
	$\PP_\mathrm{r}(t)=\|\beta^\mathrm{r}(t)\|^2$ for all $t\geq 0$,
	where $\beta^\mathrm{r}(t)$ satisfies the recursion
	\begin{align}\label{eq:betat}
		\beta^\mathrm{r}(t+1)=\Big(\big(\mathpzc{x}^\sharp_{\mathrm{k}}(t+1)\big)^\top\otimes I^{\otimes(n-k)}\Big) U_t 
		\Big(\mathpzc{x}^\sharp_{\mathrm{k}}(t)\otimes I^{\otimes(n-k)}\Big) \frac{\beta^\mathrm{r}(t)}{\|\beta^\mathrm{r}(t)\|}
	\end{align}
	with $\beta^\mathrm{r}_i(0)=a_{(\lfloor{\mathpzc{x}}_{\mathrm{k}}(0)\rfloor-1)2^{n-k}+i}$, $i=1,\dots,2^{n-k}$.
\end{theorem}

 \bluetext{The fact that with local measurements the induced Boolean dynamics becomes non-Markovian is indeed quite natural. The dark qubits carry out information that is needed for determining the full state-transition, whose evolution in turn depends on the entire history.} Note that to calculate $\PP_\mathrm{r}(t+1)$ from basic quantum measurement mechanism, one needs to record the entire path history $\mathpzc{x}_{\mathrm{k}}(0),\dots,\mathpzc{x}_{\mathrm{k}}(t+1)$. While the computing process from  Theorem \ref{thm:local.measurement} is recursive as from $\PP_\mathrm{r}(t)$ to $\PP_\mathrm{r}(t+1)$ we only need $\mathpzc{x}_{\mathrm{k}}(t)$, $\mathpzc{x}_{\mathrm{k}}(t+1)$, and $\beta^\mathrm{r}(t)$. 
The proof of Theorem~\ref{thm:local.measurement} can be found in the Appendix.

The following example is an illustration of the computation for non-Markovian transition probabilities. 

\begin{example}\label{ex:lm}
	\normalfont
	We consider a three-qubit network. Let a local measurement be $\M\otimes \M\otimes \mathrm{I}$ over qubits $1$ and $2$. Then the set of possible measurement outcomes is $\{0,1\}^2$. Let the unitary operator resulting from the continuous evolution be 
	\begin{align}
		\U_t\equiv\U=(\braket{0}{1}+\braket{1}{0})\otimes\Big(\frac{\sqrt{3}}{2}\braket{0}{0}+\frac{1}{2}\braket{0}{1}-
		\frac{1}{2}\braket{1}{0}+\frac{\sqrt{3}}{2}\braket{1}{1}\Big)\otimes(\braket{0}{0}-\braket{1}{1}).
	\end{align}
	Let the network initial state be given by
	$$
	\bra{\psi(0)}=\frac{1}{\sqrt{2}}\bra{000}+\frac{1}{\sqrt{6}}\bra{010}+\frac{1}{2\sqrt{3}}\bra{011}+\frac{1}{2}\bra{101}.
	$$ 
	Let a sample path of $\mathpzc{x}_{\mathrm{k}}(t)$ for $t=0,1,2,3$ be given by
	$$
	\mathpzc{x}_{\mathrm{k}}(0)=10,~~
	\mathpzc{x}_{\mathrm{k}}(1)=00,~~
	\mathpzc{x}_{\mathrm{k}}(2)=11,~~ 
	\mathpzc{x}_{\mathrm{k}}(3)=10. 
	$$
	From the quantum state evolution one can directly verify that
	$$
	\begin{aligned}
	\PP_\mathrm{r}(0)&=\mathbb{P}(\mathpzc{x}_{\mathrm{k}}(0)=10)=\frac{1}{4},\\
	\PP_\mathrm{r}(1)&=\mathbb{P}(\mathpzc{x}_{\mathrm{k}}(1)=00|\mathpzc{x}_{\mathrm{k}}(0)=10)=\frac{3}{4},\\
	\PP_\mathrm{r}(2)&=\mathbb{P}(\mathpzc{x}_{\mathrm{k}}(2)=11|\mathpzc{x}_{\mathrm{k}}(1)=00,\mathpzc{x}_{\mathrm{k}}(0)=10)=\frac{1}{4},\\
	\PP_\mathrm{r}(3)&=\mathbb{P}(\mathpzc{x}_{\mathrm{k}}(3)=10|\mathpzc{x}_{\mathrm{k}}(2)=11,\mathpzc{x}_{\mathrm{k}}(1)=00,\mathpzc{x}_{\mathrm{k}}(0)=10)\nonumber\\
	&=\frac{3}{4}.\\
	\end{aligned}
	$$
	Alternatively, from the recursion \eqref{eq:betat} one has
	\begin{align*}
		& \beta^\mathrm{r}(0)=\left(0,\frac{1}{2}\right)^\top,\\
		& \beta^\mathrm{r}(1)=\left(0,\frac{\sqrt{3}}{2}\right)^\top,\\
		& \beta^\mathrm{r}(2)=\left(0,-\frac{1}{2}\right)^\top,\\
		& \beta^\mathrm{r}(3)=\left(0,-\frac{\sqrt{3}}{2}\right)^\top.
	\end{align*}
	We can easily verify $\PP_\mathrm{r}(t)=\|\beta^\mathrm{r}(t)\|^2$ for $t=0,1,2,3$. This validates Theorem \ref{thm:local.measurement}.
\end{example}

\section{Controllability Conditions}\label{seccontrollability}
 \bluetext{The controllability of the quantum states under the bilinear model described by (\ref{eq:hybrid.network}) has been well understood \cite{alb02}. However, it is unclear how the random jumping in (\ref{eq:right.state}) from  the sequential measurements affects the controllability of the quantum states, or how the quantum state controllability determines the controllability of the induced Boolean dynamical states. This section   attempts to provide clear answers to these two questions.} 
\subsection{Quantum State Controllability}
It is natural to define the quantum network state controllability over the discrete state sequence $\bra{\psi(t)}=\bra{q((tT)^-)}$, $t=0,1,2,\dots$.  Note that, the sequence  $\bra{\psi(t)},t=0,1,2,\dots$ along the system
(\ref{eq:hybrid.network})--(\ref{eq:right.state}) defines a random process in its own right  as the randomness in the $\bra{\psi(t)}_\rmp$ will be inherited by $\bra{\psi(t+1)}$ for any $t$. The classical definition of the controllability of bilinear quantum systems therefore needs to be refined to accommodate  the existence of the measurements.  

We introduce the following definition of controllability for the hybrid bilinear quantum system (\ref{eq:hybrid.network})--(\ref{eq:right.state}). 

\begin{definition}\label{def:quantum.network.controllability}
	The  quantum network (\ref{eq:hybrid.network})--(\ref{eq:right.state}) is quantum state controllable if for any pair of network states $\bra{\psi_0},~\bra{\psi_1}\in\HH^{\otimes n}$, there exist an integer $T_0>0$, a global measurement $\M^{\otimes n}$, and control signals $u_\ell(s), s\in [0,T_0T]$ that steer the state of the quantum hybrid network from $\bra{\psi(0)}=\bra{\psi_0}$ to $\bra{\psi(T_0)}=\bra{\psi_1}$ with probability one.
\end{definition}

Here steering  the state of the quantum  network from $\bra{\psi(0)}=\bra{\psi_0}$ to $\bra{\psi(T_0)}=\bra{\psi_1}$ deterministically means the event that $\bra{\psi(T_0)}=\bra{\psi_1}$ conditioned that $\bra{\psi(0)}=\bra{\psi_0}$ is a sure event along (\ref{eq:hybrid.network})--(\ref{eq:right.state}). If the control signals $u_\ell(s)$, $\ell=1,\dots,p$ are feedforward, there exist deterministic unitary operators $\mathsf{U}_t$ for $t=0,1,2,\dots$ such that $\mathsf{U}_t\bra{\psi(t)}_\rmp=\bra{\psi(t+1)}$. Clearly, in this case, it is possible for the sequence  $\bra{\psi(t)},t=0,1,2,\dots$ to have degenerate probability distribution taking one possible path, but only for specially selected 
$\bra{\psi(0)},t=0,1,2,\dots$, $\mathsf{M}$, and $u_\ell(s)$, $\ell=1,\dots,p$. In particular, for that probabilistically degenerate path to take place $\bra{\psi(t)}$ must be one of the eigenvectors of the measurement $\mathsf{M}^{\otimes n}$.   As a result, the above deterministic  quantum state controllability can only be achieved by feedback controllers.  We present the following result.

\begin{proposition} \label{prop1} 
	Let $\oH_0=0$.
	Fix an arbitrary global measurement  $\M^{\otimes n}$. Then for any $T>0$, the   quantum network (\ref{eq:hybrid.network})--(\ref{eq:right.state})  is quantum state  controllable if and only if $\mathcal{L}\{\B_1,\dots,\B_p\}$ is isomorphic to $\spla(2^{n-1})$ or $\su(2^n)$.
\end{proposition}

When the network dynamics contains uncontrolled drift item, the analysis becomes more involved and we introduce  the following definition.

\begin{definition}\label{def:quantum.network.esc}
	The quantum network (\ref{eq:hybrid.network})--(\ref{eq:right.state}) is Quantum  Equivalent State Controllable if for any pair quantum states $\bra{\psi_0},\bra{\psi_1}\in\HH^{\otimes n}$, there exist an integer $T_0>0$, a global measurement $\M^{\otimes n}$, control signals $u_\ell(s), s\in [0,T_0T]$, and a phase factor $\phi$ that steer the state of the quantum network from $\bra{\psi(0)}=\bra{\psi_0}$ to $\bra{\psi(T_0)}=e^{\imath\phi}\bra{\psi_1}$ deterministically.
\end{definition}

We recall the following definition introduced  in \citeasnoun{jur72}:
$$
\begin{aligned}
R(\mathrm{I},s)=\Big\{\U\in e^{\mathcal{L}\{\A,\B_1,\dots,\B_p\}}:&\U=\X(s)\text{ is the solution}
\text{of \eqref{eq:propogator-final} under some controls $u_\ell(\cdot)$} \\ &\text{(or is reachable}
\text{along \eqref{eq:propogator-final}) at time $s$ from $\mathrm{I}$} \Big\}.
\end{aligned}
$$
We also define  $\mathbf{R}(\mathrm{I},T) = \bigcup_{0\le s\le T}R(\mathrm{I},s)$. 
\begin{proposition}\label{propc2}
	Suppose $\mathsf{A}=\imath\oH_0^{\otimes n}$ for some $\oH_0\in\mathcal{L}_*(\HH)$.
	The quantum network (\ref{eq:hybrid.network})--(\ref{eq:right.state})  is quantum  equivalent state controllable if the following conditions hold:
	
	(i) $\mathcal{L}\{\A,\B_1,\dots,\B_p\}$ is isomorphic to $\spla(2^{n-1})$ or $\su(2^n)$; 
	
	(ii) $T$ is sufficiently large so that $\mathbf{R}(\mathrm{I},T)=e^{\mathcal{L}\{\A,\B_1,\dots,\B_p\}}$.
\end{proposition}

\subsection{Boolean State Controllability}
We can also define the controllability on the induced Boolean network dynamics $\{\mathbf{x}(t)\}_{t=0}^\infty$. 

\begin{definition}\label{def:boolean.network.controllability}
	Let    a global network  measurement be given as $\M^{\otimes n}$.	
	The quantum network (\ref{eq:hybrid.network})--(\ref{eq:right.state}) is   almost surely Boolean  controllable if for any pair $X_0,X_1\in \{0,1\}^{\otimes n}$, there exist an integer $T_0>0$, and control signals $u_\ell(s), s\in [0,T_0T]$ that steer the state of the random Boolean network from ${\mathbf{x}(0)}={X_0}$ to $\mathbf{x}(T_0)={X_1}$ with probability one along the induced Boolean dynamics $\{\mathbf{x}(t)\}_{t=0}^\infty$.
\end{definition}

 \bluetext{
	It is straightforward to verify that  Boolean controllability is an inherently relaxed  controllability notion. We introduce the following definition of practical controllability of the quantum states concerning whether controllability can be achieved in the approximate sense \cite{practical}. 
	
	\begin{definition}
		The bilinear control system (\ref{eq:hybrid.network})
		is practically controllable with respective to $\delta >0$ if for any $\bra{\psi_0}$ and $\bra{\psi_f}$ there exist $u_\ell(s):s\in[0,T), \ell=1,\dots,p$ such that
		$$\bra{q(0)} = \bra{\psi_0}
		\Longrightarrow
		\big\| \bra{q(T)} - \bra{\psi_f}\big\|<\delta
		$$
	\end{definition}
	
	We now present the following result suggesting that practical controllability for the quantum states implies almost sure controllability for the induced Boolean states. 
	\begin{theorem}\label{thm4}
		Let the   bilinear control system (\ref{eq:hybrid.network})
		be practically controllable with respective to some $\delta$ with $\delta < \sqrt{2}$.
		Then 
		\begin{itemize}
			\item[(i)] The hybrid qubit network  (\ref{eq:hybrid.network})--(\ref{eq:right.state})  is almost surely Boolean controllable.
			
			\item[(ii)] For any $X_0,X_*\in \{0,1\}^{\otimes n}$,  for 
			$$T_{\rm hit} = \inf_{t\geq 0} \{X(t) = X_{*}  \}$$
			with $X(0)=X_0$ there holds 
			\begin{align*}
				\max_{u_\ell(s):s\in[0,tT)}
				\mathbb{P} 
				(T_{\rm hit} \leq t  )
				>1-e^{-t \log \left(\frac{4}{4\delta^2-\delta^4}\right)}.
			\end{align*}
		\end{itemize}
	\end{theorem}
	
	Theorem \ref{thm4}.(ii) shows that in the presence of practical quantum state controllability, the probability of arriving at any measurement outcome $X_\ast \in\{0,1\}^{\otimes n}$ approaches one at an exponential rate. Moreover, the measurement outcome $\mathbf{x}(t)=X_\ast$ corresponds uniquely to the quantum state $|q(tT)\rangle =|X_\ast\rangle$. Therefore, this Boolean state controllability also provides a way of realizing verifiable quantum state manipulation by the combination of Bilinear control and sequential measurements.   
	The proofs of Proposition~\ref{prop1}, Proposition~\ref{propc2}, and Theorem~\ref{thm4} are in the Appendix.
}

\subsection{Further Discussions}
The  controllability  definition of the hybrid bilinear quantum network under local measurement can be similarly introduced.  
 \bluetext{
	For any initial $\bra{\psi_0}$, after being measured its post-measurement state $\bra{\psi_0}_\rmp$  is in $\{\bra{0},\bra{1}\}^{\otimes n}$, which is known even when  $\bra{\psi_0}$ is unknown. Therefore, an advantage in terms of controllability from global measurement is the fact that the initial quantum state can be uncertain for reaching any target state. However, with local measurements, the initial state   $\bra{\psi_0}$ must be fully known in order to establish any post-measurement state initial $\bra{\psi(t)}_{\rm p}$, which is critical for the design of any feedback controller. 
	This point represents the most significant difference between these two types of measurements for  the controllability properties. When the initial state $\bra{\psi_0}$ is known, similar  results can be established along the same line of analysis for the controllability of the quantum network with local measurements. 
}

It is certainly of interest to investigate how the graphical network structure influences the controllability  of the quantum networks.  \bluetext{The network structure can be defined by the drift Hamiltonian $\mathsf{H}_0$, or controlled Hamiltonians $\mathsf{H}_\ell$, where edges arise from the qubit interactions encoded in $\mathsf{H}_0$ or $\mathsf{H}_\ell$. Alternatively, generalized network structures can be defined over the interaction relationship among the $2^n$ quantum states. Excellent results have been established regarding how such an interaction structure would lead to the Lie-algebra  controllability condition   \cite{Altafini2002,li2017,belabbas2018}. We note that such results can be applied to the hybrid network model considered in the current paper as well, since the controllability in the presence of measurements is still closely related to the original bilinear  controllability as shown in the results.  }

\section{Conclusions}\label{seccon}
We have studied  dynamical quantum networks subject to sequential local or global measurements leading to  probabilistic Boolean recursions  which represent  the   measurement outcomes. With global  measurements,  such resulting Boolean recursions   were shown to be Markovian, while   with local measurements, the state transition probability at any given time depends on the entire history of the sample path.  Under the   bilinear control  model for the Schr\"odinger evolution, we showed that the measurements in general enhance the controllability of the quantum networks. The global or local  measurements were assumed to be prescribed in the current framework. It is of interest as a future direction to investigate the co-design of the continuous control signals and the measurements, which may both have local structures,  for more robust and efficient methods of manipulating the states of  large-scale quantum networks.

\section*{Appendix}

\subsection*{A. Proof of Theorem \ref{thm:1} }

From the definition of $f_{[\alpha_1,\dots,\alpha_{2^n}]}$, $F_t$ taking value as $f_{[\alpha_1,\dots,\alpha_{2^n}]}$ is equivalent to obtaining outcomes $\delta_{2^n}^1,\dots,\delta_{2^n}^{2^n}$, respectively, when measuring quantum states independently prepared at $\delta_{2^n}^1,\dots,\delta_{2^n}^{2^n}$.
Then the probability of $F_t:\mathbf{x}^\sharp(t)\rightarrow \mathbf{x}^\sharp(t+1)$ taking $f_{[\alpha_1,\dots,\alpha_{2^n}]}$ as the transition matrix is
$$
\begin{aligned}
p(f_{[\alpha_1,\dots,\alpha_{2^n}]})&=\prod\limits_{i=1}^{2^n}\mathbb{P}\left(\mathbf{x}^\sharp(t+1)=\delta_{2^n}^{\alpha_i}\bigg|\mathbf{x}^\sharp(t)=\delta_{2^n}^{i}\right).\\
\end{aligned}
$$
To express this probability, we need to figure out each $\mathbb{P}\left(\mathbf{x}^\sharp(t+1)=\delta_{2^n}^{\alpha_i}\bigg|\mathbf{x}^\sharp(t)=\delta_{2^n}^{i}\right)$.
At time $t$, if the outcome is  $[\lambda_{x_1(t)},\dots,\lambda_{x_n(t)}]\sim \mathbf{x}^\sharp(t)$, $x_j(t)\in\{0,1\},j=1,\dots,n$ after the network state $\bra{\psi(t)}$ being measured, then the probability of getting outcome $[\lambda_{x_1(t+1)},\dots,\lambda_{x_n(t+1)}]\sim \mathbf{x}^\sharp(t+1)$ is
$$
\begin{aligned}
&~~~~\mathbb{P}\Big(\mathbf{x}^\sharp(t+1)\Big|\mathbf{x}^\sharp(t)\Big)\\
&=\Big|\ket{v_{x_1(t+1)}\cdots v_{x_n(t+1)}}\U_t\bra{{v_{x_1(t)}\cdots v_{x_n(t)}}}\Big|^2\\
&=\Big|\ket{{x_1(t+1)}\cdots {x_n(t+1)}}(\underbrace{\uu\otimes\cdots\otimes \uu}_n)^\dag 
\U_t\underbrace{\uu\otimes\cdots\otimes \uu}_n\bra{{{x_1(t)}\cdots {x_n(t)}}}\Big|^2\\
&=\Big|[U_t^\M]_{\lfloor\mathbf{x}(t+1)\rfloor,\lfloor\mathbf{x}(t)\rfloor}\Big|^2.
\end{aligned}
$$

Since $\lfloor\mathbf{x}(t)\rfloor,\lfloor\mathbf{x}(t+1)\rfloor\in\{1,2,\dots,2^n\}$, we have
$$
\mathbb{P}\left(\mathbf{x}^\sharp(t+1)=\delta_{2^n}^{\alpha_i}\bigg|\mathbf{x}^\sharp(t)=\delta_{2^n}^{i}\right)=\Big| [U_t^\M]_{\alpha_i,i}\Big|^2.
$$
Thus, the probability of $F_t$ taking $f_{[\alpha_1,\dots,\alpha_{2^n}]}$ is
$$
\begin{aligned}
p(f_{[\alpha_1,\dots,\alpha_{2^n}]})&=\prod\limits_{i=1}^{2^n}\mathbb{P}\left(\mathbf{x}^\sharp(t+1)=\delta_{2^n}^{\alpha_i}\bigg|\mathbf{x}^\sharp(t)=\delta_{2^n}^{i}\right)\\
&=\prod\limits_{i=1}^{2^n} \Big|[U_t^\M]_{\alpha_i,i}\Big|^2.
\end{aligned}
$$
This completes the proof.  

\subsection*{B. Proof of Theorem \ref{thm:local.measurement}}

We first  present the following technical lemma on the tensor product of projector matrices, which can be verified directly.  
\begin{lemma}\label{lem:2}
	Denote \begin{align}\label{eq100}
		P_0=\begin{pmatrix}1&0\\0&0\end{pmatrix},
		P_1=\begin{pmatrix}0&0\\0&1\end{pmatrix}.
	\end{align} 
	Let $\gamma=[\gamma_1,\dots,\gamma_k]$, where $\gamma_i\in\{0,1\},~i=1,\dots,k$. Then
	$$
	[P_{\gamma_1}\otimes\dots\otimes P_{\gamma_k}]_{i,j} = 
	\begin{cases}
	1, & i=j=\lfloor\gamma\rfloor,\\
	0, & \text{otherwise}.
	\end{cases}
	$$
\end{lemma}

First, if we measure $\bra{\psi(0)}$ and get outcome $\mathpzc{x}_{\mathrm{k}}(0)$, then the probability of getting $\mathpzc{x}_{\mathrm{k}}(0)$ is
$$
\begin{aligned}
\PP_\mathrm{r}(0) & = \ket{\psi(0)}\P_{x_1(0)}\otimes\dots\otimes \P_{x_k(0)}\otimes \mathrm{I}^{\otimes(n-k)}\bra{\psi(0)} \\
& = \sum_{i=1}^{2^{n-k}}|a_{(\lfloor{\mathpzc{x}}_{\mathrm{k}}(0)\rfloor-1)2^{n-k}+i}|^2\\
&= \|\beta^\mathrm{r}(0)\|^2
\end{aligned}
$$
with $\beta^\mathrm{r}_i(0)=a_{(\lfloor{\mathpzc{x}}_{\mathrm{k}}(0)\rfloor-1)2^{n-k}+i}$, $i=1,\dots,2^{n-k}$.
Moreover, given $\mathpzc{x}_{\mathrm{k}}(0)$ occured, the vector form of the post-measurement state of $\bra{\psi(0)}$ under the chosen basis is
$$
\begin{aligned}
\bra{\psi(0)}_\rmp^\mathrm{r} & = \frac{\P_{x_1(0)}\otimes\dots \otimes\P_{x_k(0)}\otimes \mathrm{I}^{\otimes(n-k)}\bra{\psi(0)}}{\sqrt{\|\beta^\mathrm{r}(0)\|^2}}\\
&=\frac{\sum_{i=1}^{2^{n-k}}\beta^\mathrm{r}_i(0)\bra{b_{(\lfloor{\mathpzc{x}}_{\mathrm{k}}(0)\rfloor-1)2^{n-k}+i}}}{\|\beta^\mathrm{r}(0)\|},\\
&=\bra{x_1(0)\cdots x_k(0)}\otimes\frac{\sum_{i=1}^{2^{n-k}}\beta^\mathrm{r}_i(0)\bra{b_{i}}^{(n-k)}}{\|\beta^\mathrm{r}(0)\|},
\end{aligned}
$$
where Lemma~\ref{lem:2} is used in the second equality, and $\{\bra{b_{i}}^{(n-k)},i=1,\dots,2^{n-k}\}=\{\bra{0},\bra{1}\}^{\otimes (n-k)}$.

Next, we compute $\PP_\mathrm{r}(1)$. Given $\mathpzc{x}_{\mathrm{k}}(0)$, the network state at time $1$ is
$$
\bra{\psi(1)}^\mathrm{r}=\U_0\bra{\psi(0)}_\rmp^\mathrm{r}.
$$
Subject to $\M^{\mathrm{V}_*}=\M^{\otimes k}\otimes \mathrm{I}^{\otimes(n-k)}$, the probability of getting outcome $\mathpzc{x}_{\mathrm{k}}(1)$ is
$$
\begin{aligned}
&\PP_\mathrm{r}(1) = \ket{\psi(1)}\P_{x_1(1)}\otimes\dots\otimes \P_{x_k(1)}\otimes \mathrm{I}^{\otimes(n-k)}\bra{\psi(1)} \\
&= \sum_{i=1}^{2^{n-k}}\left|\frac{\sum_{j=1}^{2^{n-k}}\beta^\mathrm{r}_i(0)[U_0]_{(\lfloor{\mathpzc{x}}_{\mathrm{k}}(1)\rfloor-1)2^{n-k}+i,(\lfloor{\mathpzc{x}}_{\mathrm{k}}(0)\rfloor-1)2^{n-k}+j}}{\|\beta^\mathrm{r}(0)\|}\right|^2 \\
&= \|\beta^\mathrm{r}(1)\|^2
\end{aligned}
$$
with
\begin{align}\label{eq:beta1}
	\beta^\mathrm{r}_i(1)
	&=\frac{\sum_{j=1}^{2^{n-k}}\beta^\mathrm{r}_i(0)[U_0]_{(\lfloor{\mathpzc{x}}_{\mathrm{k}}(1)\rfloor-1)2^{n-k}+i,(\lfloor{\mathpzc{x}}_{\mathrm{k}}(0)\rfloor-1)2^{n-k}+j}}{\|\beta^\mathrm{r}(0)\|}\nonumber\\
	&=\frac{\Big(\big(\mathpzc{x}^\sharp_{\mathrm{k}}(1)\big)^\top\otimes I^{\otimes(n-k)}\Big) U_0 \Big(\mathpzc{x}^\sharp_{\mathrm{k}}(0)\otimes I^{\otimes(n-k)}\Big) \beta^\mathrm{r}(0)}{\|\beta^\mathrm{r}(0)\|},
\end{align}
for $i=1,\dots,2^{n-k}$.
Similarly, given $\mathpzc{x}_{\mathrm{k}}(0)$ and $\mathpzc{x}_{\mathrm{k}}(1)$, the vector form of post-measurement state of $\bra{\psi(1)}^\mathrm{r}$ depending on $\mathrm{r}$ is
$$
\begin{aligned}
\bra{\psi(1)}_\rmp^\mathrm{r} & = \frac{\P_{x_1(1)}\otimes\dots\otimes \P_{x_k(1)}\otimes \mathrm{I}^{\otimes(n-k)}\bra{\psi(1)}}{\sqrt{\|\beta^\mathrm{r}(1)\|^2}}\\
&=\frac{\sum_{i=1}^{2^{n-k}}\beta^\mathrm{r}_i(1)\bra{b_{(\lfloor{\mathpzc{x}}_{\mathrm{k}}(1)\rfloor-1)2^{n-k}+i}}}{\|\beta^\mathrm{r}(1)\|},\\
&=\bra{x_1(1)\cdots x_k(1)}\otimes\frac{\sum_{i=1}^{2^{n-k}}\beta^\mathrm{r}_i(1)\bra{b_{i}}^{(n-k)}}{\|\beta^\mathrm{r}(1)\|}.
\end{aligned}
$$

Finally, the above process can be carried out recursively, so that $\PP_\mathrm{r}(2),~\PP_\mathrm{r}(3),~\dots,~\PP_\mathrm{r}(t+1)$ can be computed from this procedure. The recursion from $\PP_\mathrm{r}(i)$ to $\PP_\mathrm{r}(i+1)$, $i\ge 1$ will follow from the same process as $\PP_\mathrm{r}(0)$ to $\PP_\mathrm{r}(1)$, and we can establish \eqref{eq:betat} eventually.

This completes the proof. 

\subsection*{C. Proof of Proposition \ref{prop1}}

With feedback controllers, it is clear from the Markovian property of $\mathsf{U}_t\bra{\psi(t)}_\rmp$ that 
we can assume $T_0=1$ for the definition of the quantum state controllability.   \bluetext{After the measurement at $t=0$, the post-measurement state $\bra{\psi_0}_\rmp$ of any initial state $\bra{\psi_0}$ belongs to $\{\bra{0},\bra{1}\}^{\otimes n}$ which is a finite set but  is still a subset of $\QQ(2^n)$. 
	The sufficiency statement is therefore a special case 
	of classical result, e.g., Theorem 5 in \citeasnoun{jur72}.  
	
	Now, we prove  the necessity continues to hold.  Suppose the quantum network is quantum state controllable. Then with $\bra{\psi_0}\in\{\bra{0},\bra{1}\}^{\otimes n}$ and  for any $\bra{\psi_1}\in \QQ(2^n)$, there exist control signal $u_\ell(s),~s\in[0,T)$, such that $\bra{\psi(0)}=\bra{\psi_0}$  and $\bra{\psi(T)}=\bra{\psi_1}$. 
	Thus there exists $\U_{\bra{\psi_1}}$ such that $\bra{\psi_1}=\U_{\bra{\psi_1}}\bra{\psi_0}$. By Theorem 5 in \citeasnoun{bro72}, the solution at $s=T$ of \eqref{eq:propogator-final} from $\mathrm{I}$ at $s=0$ is $\mathsf{X}(T)=\U_{\bra{\psi_1}}\in e^{\mathcal{L}\{\B_1,\dots,\B_p\}}$. Denoting $\mathbf{U}=\{\U_{\bra{\psi_1}}:\bra{\psi_1}\in \QQ(2^n)\}$, we have the following facts: 
	\begin{itemize}
		\item [(i)] $\mathbf{U}\subseteq e^{\mathcal{L}\{\B_1,\dots,\B_p\}}$; 
		\item [(ii)] $\mathbf{U}\bra{\psi_0}=\QQ(2^n)$; 
		\item[(iii)] $\U_\ast\bra{\psi_0}\in \QQ(2^n)$, for any $\U_\ast\in e^{\mathcal{L}\{\B_1,\dots,\B_p\}}$.
	\end{itemize} Hence  $e^{\mathcal{L}\{\B_1,\dots,\B_p\}}\bra{\psi_0}=\QQ(2^n)$. Because of the reversibility of the action of elements in the group $e^{\mathcal{L}\{\B_1,\dots,\B_p\}}$, we can further conclude that $e^{\mathcal{L}\{\B_1,\dots,\B_p\}}$ is transitive on $\QQ(2^n)$. Invoking Theorem 4 of \cite{alb03}, the desired conclusion holds. }

\subsection*{D. Proof of Proposition \ref{propc2}}

Let $\M=\oH_0$. 
For any pair quantum states $\bra{\psi_0}$ and $\bra{\psi_1}$, 
the post-measurement state of $\bra{\psi_0}$ being measured by $\mathsf{H}_0^{\otimes n}$ is $\bra{\psi_0}_\rmp$, which is an eigenstate of $\mathsf{H}_0^{\otimes n}$. We let the corresponding eigenvalue of $\bra{\psi_0}_\rmp$ is $\lambda$.
If $\mathcal{L}\{\A,\B_1,\dots,\B_p\}$ is isomorphic to $\spla(2^{n-1})$ or $\su(2^n)$, then $\mathcal{L}\{\A,\B_1,\dots,\B_p\}$ is transitive. From Theorem 6.5 of \citeasnoun{jur72} with the condition that $T$ is sufficiently large so that $\mathbf{R}(\mathrm{I},T)=e^{\mathcal{L}\{\A,\B_1,\dots,\B_p\}}$, there exists $s_*$ such that we can find a $\U\in R(\mathrm{I},s_*)$ with controls $u_\ell^*(\cdot)$ such that $\bra{\psi_1}=\U\bra{\psi_0}_\rmp$. Now we set the admissible control as
$$
u_\ell(s)=\begin{cases}
0, & s\in[0,T-s_*] \\
u^*_\ell(s), & s\in(T-s_*,T]
\end{cases},\quad \ell=1,\dots,p.
$$
Under this control, the system state $\bra{\varphi(s)}$ will be driven to (1) $
e^{-\imath\lambda}\bra{\psi_0}_\rmp$ at time $s=T-s_*$ from $\bra{\psi_0}_\rmp$; (2) $
e^{-\imath\lambda}\bra{\psi_1}$ at time $s=T$.
This completes the proof.   

 \bluetext{
	\subsection*{E. Proof of Theorem \ref{thm4}}
	(i) Denote the quantum state corresponding to the measurement outcome $X_0,X_*\in\{0,1\}^{\otimes n}$ as $|X_0\rangle$ and $|X_*\rangle$, respectively. Since   bilinear control system (\ref{eq:hybrid.network})
	is practically controllable with respective to some $\delta$ with $\delta < \sqrt{2}$, for any $|\psi(t)\rangle_{\rm p}=|\mathbf{x}(t)\rangle$, there always exists $u_\ell(s):s\in[tT,(t+1)T)$ such that 
	\begin{align}
		\Big| \langle X_*|\psi(t+1)\rangle \Big| &\geq  {\rm Re} \big( \langle X_*|\psi(t+1)\rangle\big) \nonumber\\
		&\geq \frac{2-\delta^2}{2}\nonumber\\
		&>0. 
	\end{align}
	As a result, there holds
	\begin{align}\label{p1}
		\mathbb{P}\Big(\mathbf{x}(t+1)=X_* \Big|\mathbf{x}(t) \Big)&\geq \Big| \langle X_*|\psi(t+1)\rangle \Big|^2\nonumber\\
		&\geq \left(1-\frac{\delta^2}{2}\right)^2
	\end{align}
	for all $t\geq 0$. The desired almost sure Boolean controllability follows directly from the Borel-Cantelli Lemma (cf. Theorem 2.3.6, \cite{durrett}).
	
	\medskip
	
	\noindent (ii) In view of (\ref{p1}) and according to the definition of $T_{\rm hit}$, there holds 
	\begin{align*}
		\max_{u_\ell(s):s\in[0,tT)}
		\mathbb{P} 
		\Big(T_{\rm hit} \geq t \Big) &\leq \left(1-\left(1-\frac{\delta^2}{2}\right)^2\right)^t\nonumber\\
		&=\left(\delta^2-\frac{\delta^4}{4}\right)^t.
	\end{align*}
	This immediately implies that 	 
	\begin{align*}
		\max_{u_\ell(s):s\in[0,tT)}
		\mathbb{P} 
		\Big(T_{\rm hit} \leq t  \Big)
		&\geq 1 - \left(\delta^2-\frac{\delta^4}{4}\right)^t\\
		& = 1-e^{-t \log \left(\frac{4}{4\delta^2-\delta^4}\right)}.
	\end{align*}
	This proves the desired theorem. }

\end{document}